\documentclass[11pt]{article}
\usepackage{arxiv}	
\usepackage{url}      
\usepackage{graphicx,times}
\usepackage{caption}
\usepackage{subcaption}
\usepackage{color}
\usepackage{natbib}
\usepackage{graphicx}
\usepackage{color}
\usepackage{hyperref}

\title{Curved Space-Filling Tiles Using Voronoi Decomposition with Line, and Curve Segments Closed Under Wallpaper Symmetries}

\author{ Haard Panchal\\
	Visualization Department, \\ Texas A\&M University, College Station, TX, 77831\\
	\texttt{haard.panchal@tamu.edu } \\
     \And
     \href{https://orcid.org/0000-0003-3618-4166}{\includegraphics[scale=0.06]{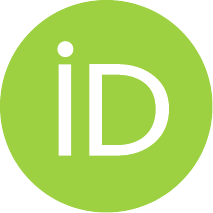}\hspace{1mm}Ergun Akleman}\thanks{Joint with Computer Science and Engineering Department.} \\
	Visual Computing \& Computational Media,\\ Texas A\&M University, College Station, TX, 77831\\
	\texttt{ergun@tamu.edu} \\
	\And
	\href{https://orcid.org/0000-0001-6434-2577}{\includegraphics[scale=0.06]{orcid.pdf}\hspace{1mm}Vinayak Krishnamurthy} \thanks{Joint with Computer Science and Engineering Department.}\\
	Department of Mechanical Engineering, \\ Texas A\&M University, College Station, TX, 77831\\
	\texttt{vinayak@tamu.edu} 
 	\And
	Tolga Talha Yildiz\\
	Computer Science and Engineering Department, \\ Texas A\&M University, College Station, TX, 77831\\
	\texttt{tolgayildiz@tamu.edu} 
     \And
     Varda Grover\\
	Visual Computing \& Computational Media,\\ Texas A\&M University, College Station, TX, 77831\\
	\texttt{varda\_grover@tamu.edu} \\
}

\renewcommand{\shorttitle}{Curved Space-Filling Tiles}

\begin{document}

\maketitle

\thispagestyle{empty}

\begin{abstract}
In this paper, we present a new approach to obtain symmetric tiles with curved edges. Our approach is based on using higher-order Voronoi sites that are closed under wallpaper symmetries. The resulting Voronoi tessellations provide us with symmetric tiles with curved edges. We have developed a web application that provides real-time tile design. Our application can be found at \href{https://voronoi.viz.tamu.edu/}{https://voronoi.viz.tamu.edu/}. One of our key findings in this paper is that not all symmetry operations are useful for creating curved tiles. In particular, all symmetries that use mirror operation produce straight lines that are useless for creating new tiles. This result is interesting because it suggests that we need to avoid mirror transformations to produce unusual space-filling tiles in 2D and 3D using Voronoi tessellations.

\end{abstract}

\begin{figure}[htpb]
    \centering
    \begin{subfigure}[t]{0.24\textwidth}
        \includegraphics[width=0.99\textwidth]{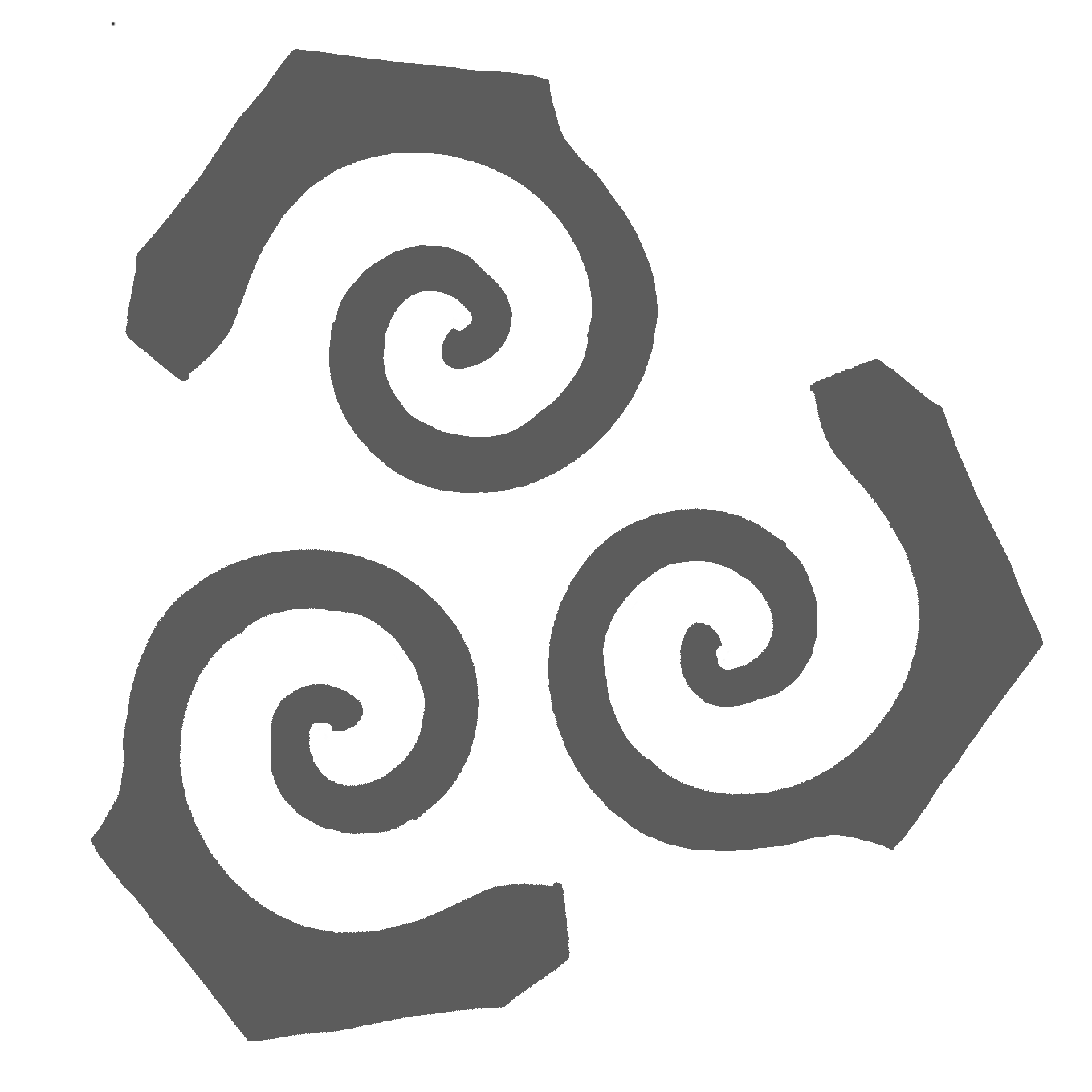}
        \caption{\it An example of curved tiles with p3 symmetry.}
        \label{renders/Escher0}
    \end{subfigure}
    \hfill
    \begin{subfigure}[t]{0.24\textwidth}
        \includegraphics[width=0.99\textwidth]{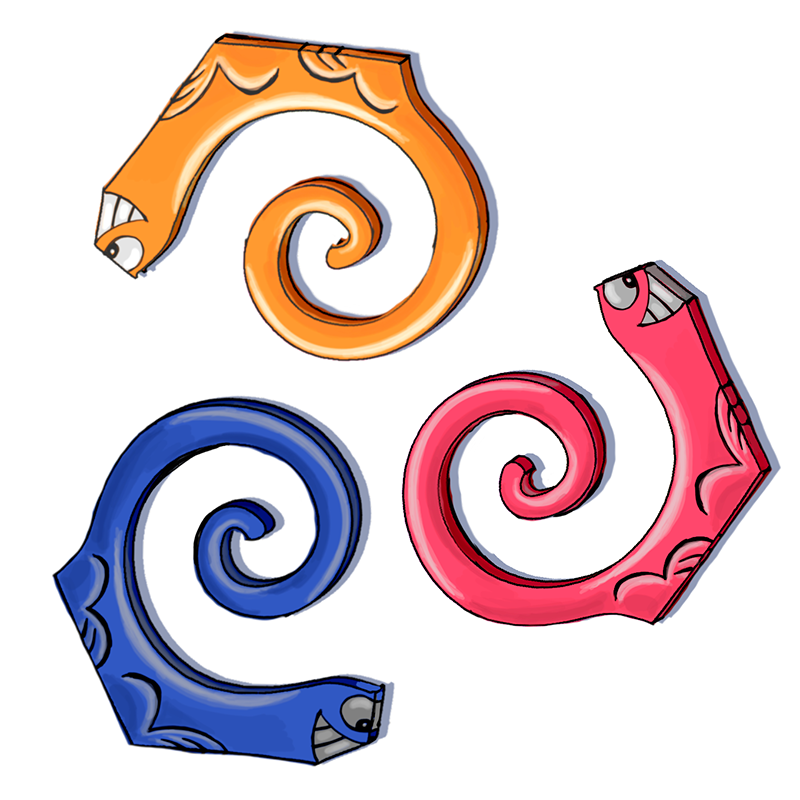}
        \caption{\it Same tiles with Escher-like hand-drawn faces.}
        \label{renders/Escher1}
    \end{subfigure}
    \hfill
    \begin{subfigure}[t]{0.24\textwidth}
        \includegraphics[width=0.99\textwidth]{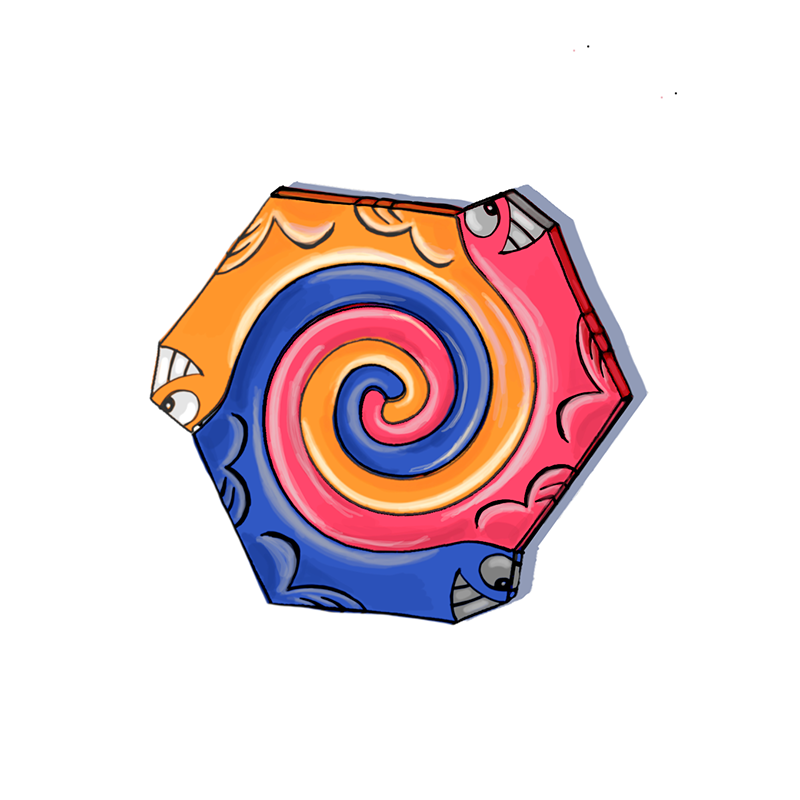}
        \caption{\it Same tiles interlocked together.}
        \label{renders/Escher2}
    \end{subfigure}
    \hfill
        \begin{subfigure}[t]{0.24\textwidth}
        \includegraphics[width=0.99\textwidth]{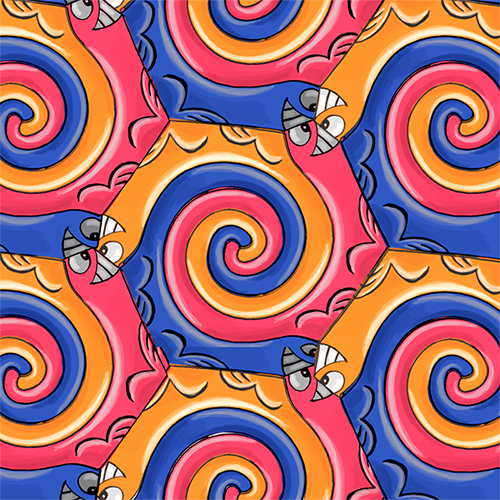}
        \caption{\it Same tiles filling 2D space.}
        \label{renders/Escher3}
    \end{subfigure}
    \hfill
\caption{\it Using our system artists can create Escher-like curved congruent tiles. In this example, we demonstrated how a congruent tile was created in our system using p3 symmetry can later be painted to obtain the look of Escher-style animal drawings that can tesselate the 2D space.}
\label{Escher_Example}
\end{figure}

\section{Introduction and Motivation}

With the development of the internet and computer graphics, many interactive web-based systems to design symmetric patterns have been developed. Most web-based symmetric pattern design systems are really painting systems, that do not produce tiles. One alternative approach is to develop them as drawing systems. There currently exist two types of drawing systems: (1) The stand-alone systems that allow for manipulating polygons based on Escher's method of creating space-filling tiles such as Tesselmaniac \cite{tessselmaniac} and Tesselation Exploration; and  (2) the web-based systems that allow drawing graphs based on 17 wallpaper symmetries, such as Kali \cite{amenta1995}, Symmetric Graph Drawing system \cite{akleman2000web}, and EscherSketch \cite{levskaya2017}. The advantage of the first approach is that it guarantees watertight tessellations without gaps and overlaps. However, it only allows for a change of the geometric shape of the polygonal tiles \cite{tessselmaniac}. 

\begin{figure}[htpb]
    \centering
    \begin{subfigure}[t]{0.48\textwidth}
        \includegraphics[width=0.48\textwidth]{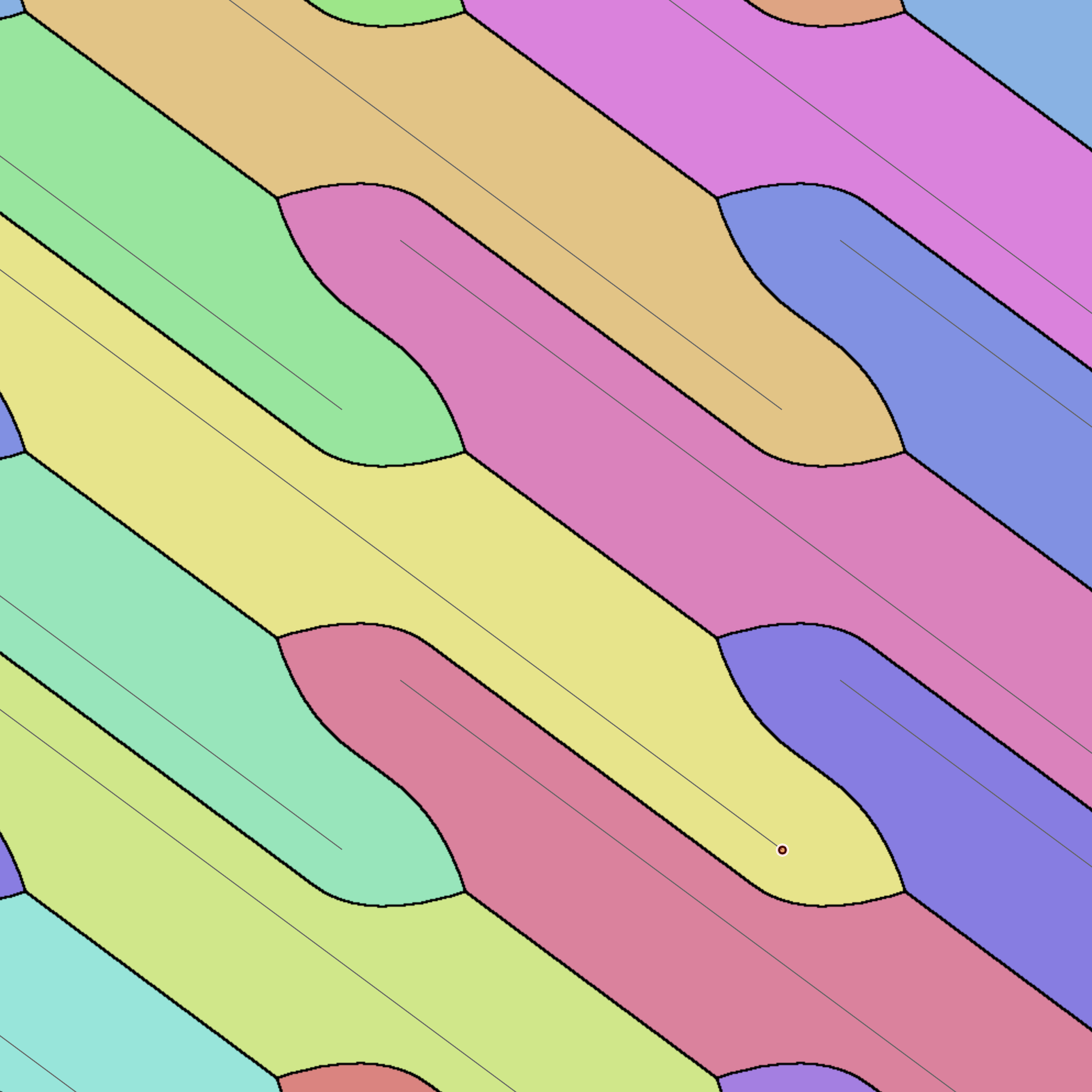}
        \includegraphics[width=0.48\textwidth]{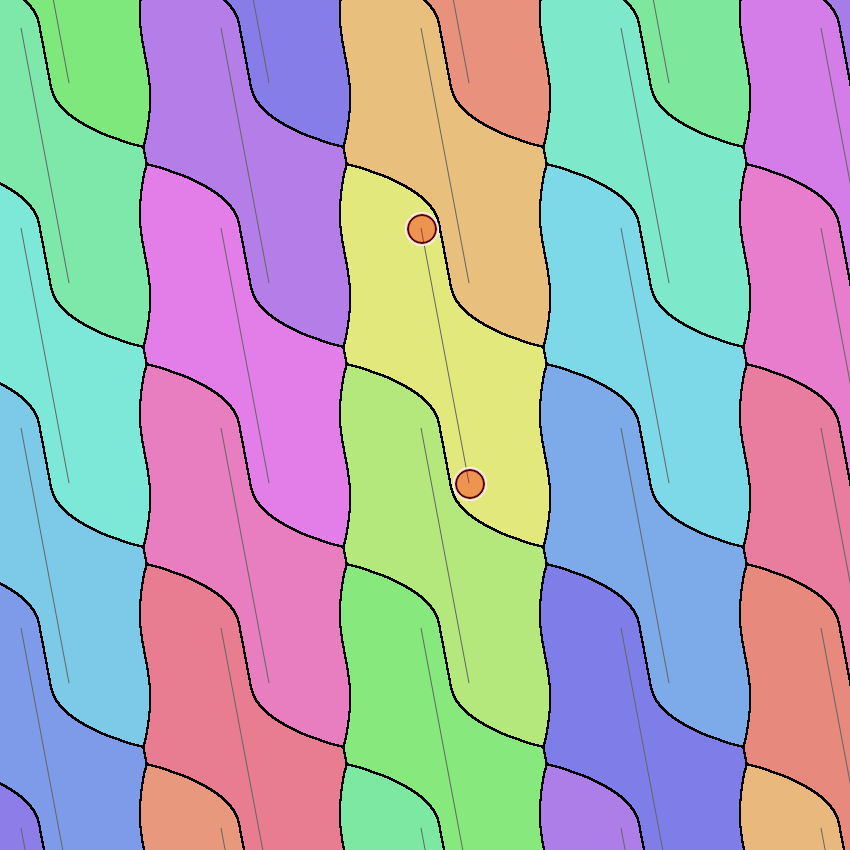}
        \caption{\it Patterns with p1 symmetry.}
        \label{lines/p1}
    \end{subfigure}
    \hfill
    \begin{subfigure}[t]{0.48\textwidth}
        \includegraphics[width=0.48\textwidth]{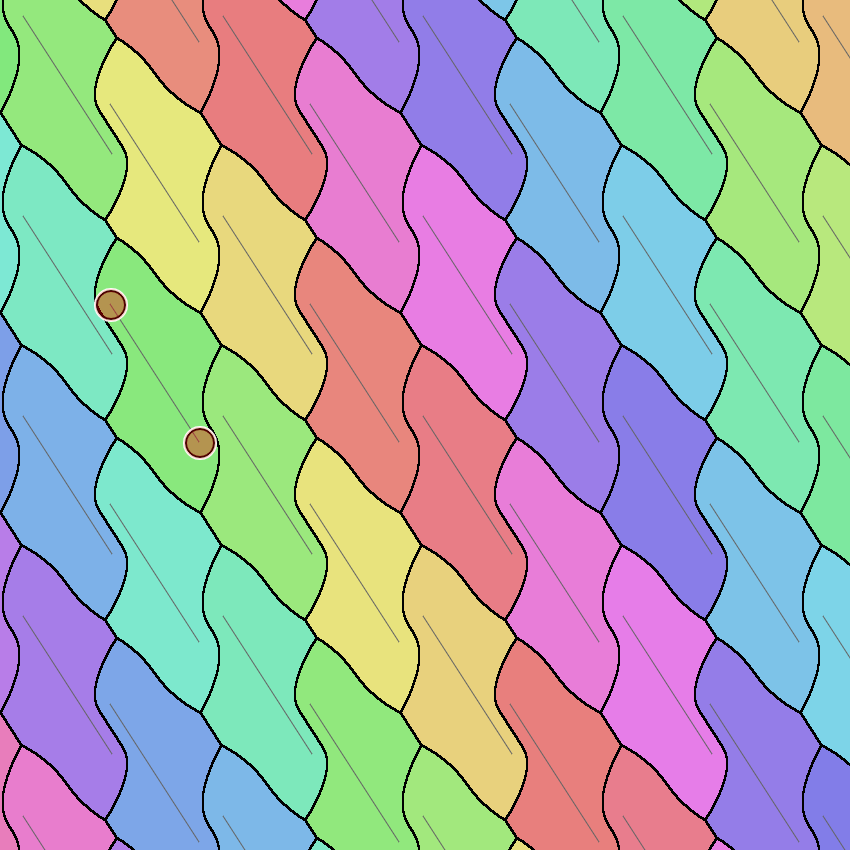}
        \includegraphics[width=0.48\textwidth]{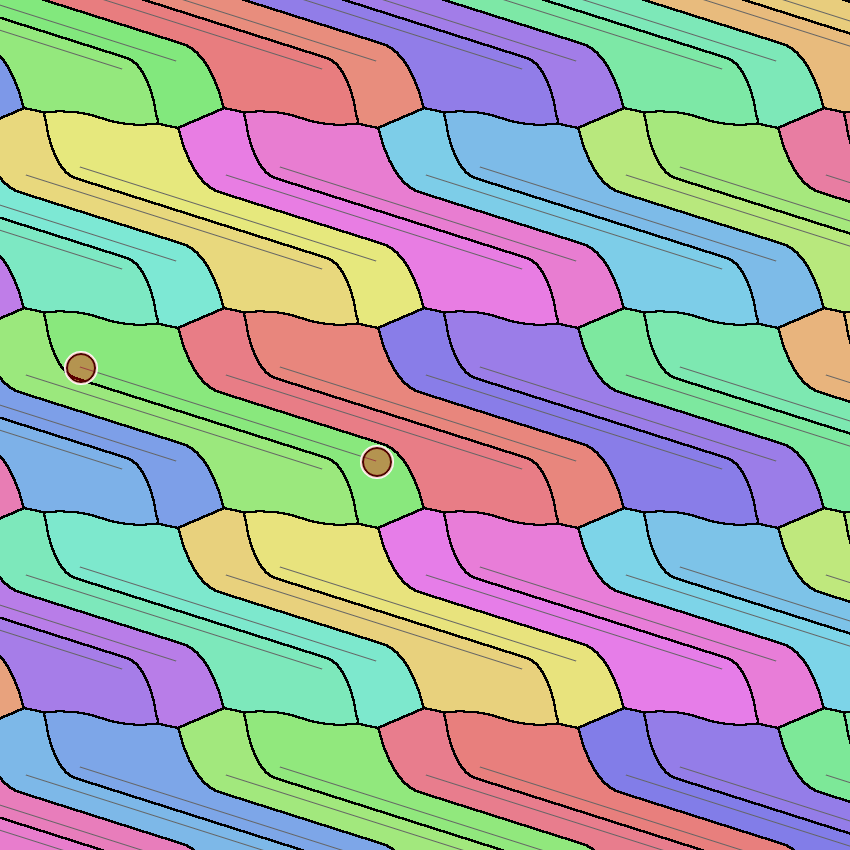}
        \caption{\it Patterns with p2 symmetry.}
        \label{lines/p2}
    \end{subfigure}
    \hfill
    \begin{subfigure}[t]{0.48\textwidth}
        \includegraphics[width=0.48\textwidth]{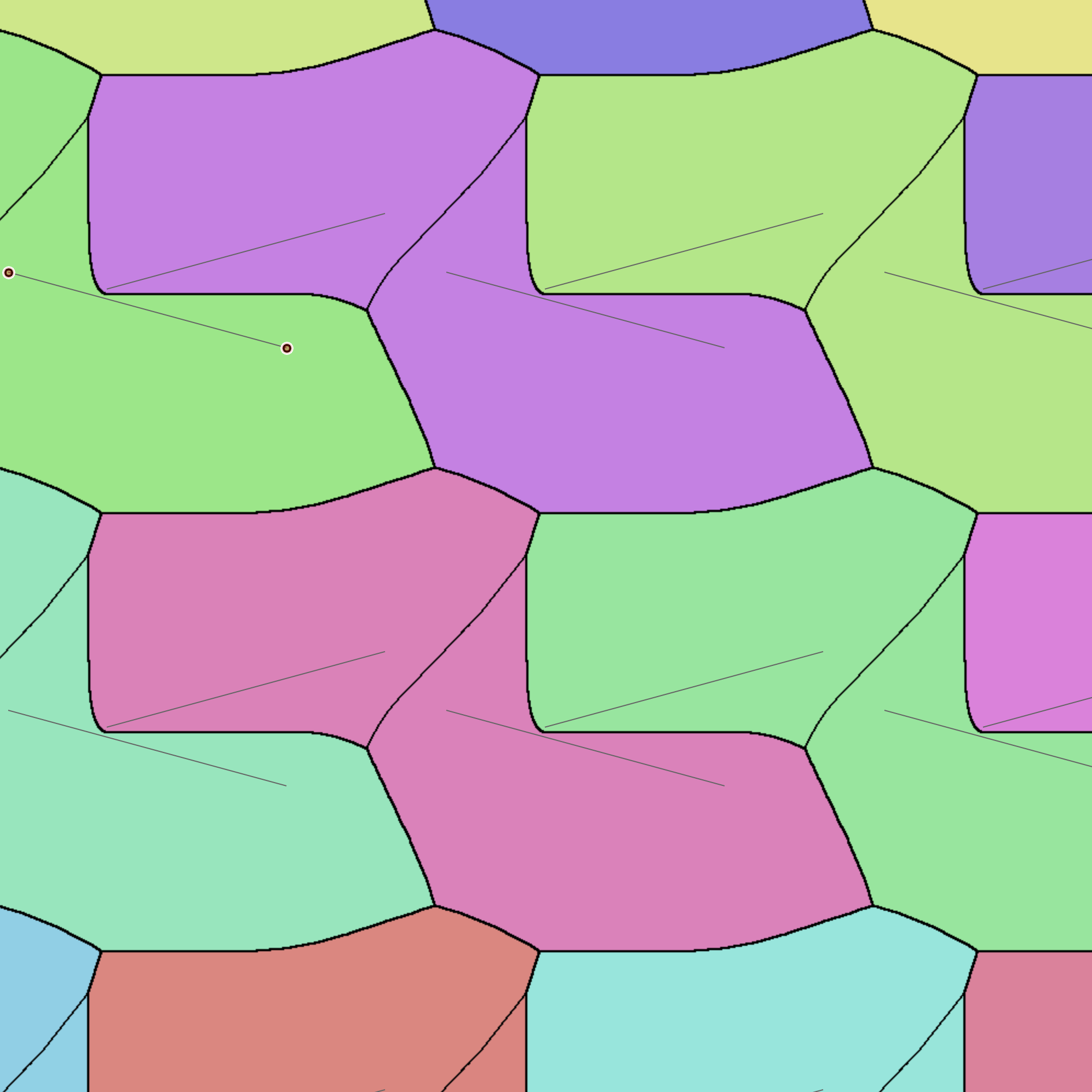}
        \includegraphics[width=0.48\textwidth]{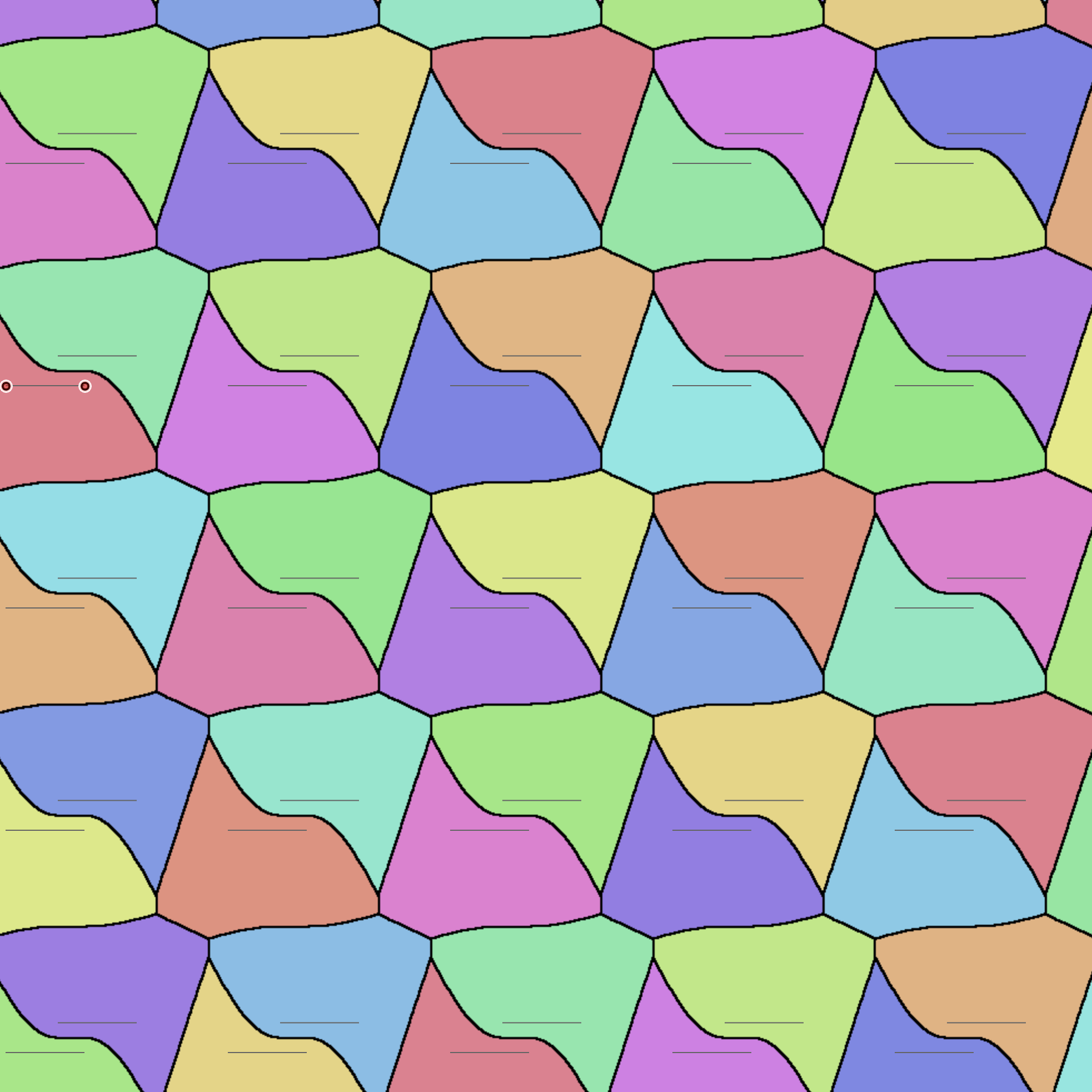}
        \caption{\it Patterns with pg symmetry.}
        \label{lines/pg_1}
    \end{subfigure}
    \hfill
    \begin{subfigure}[t]{0.48\textwidth}
        \includegraphics[width=0.48\textwidth]{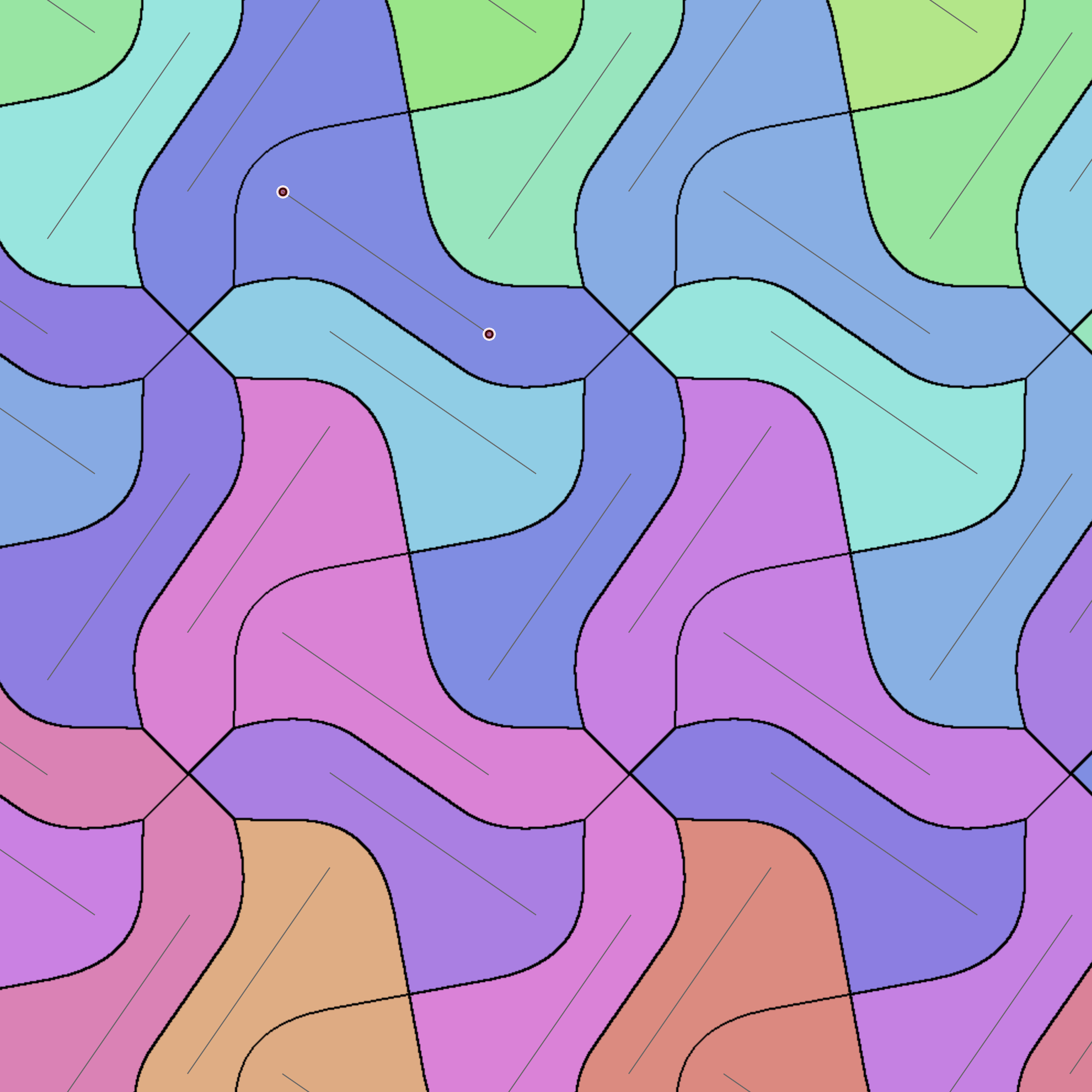}
        \includegraphics[width=0.48\textwidth]{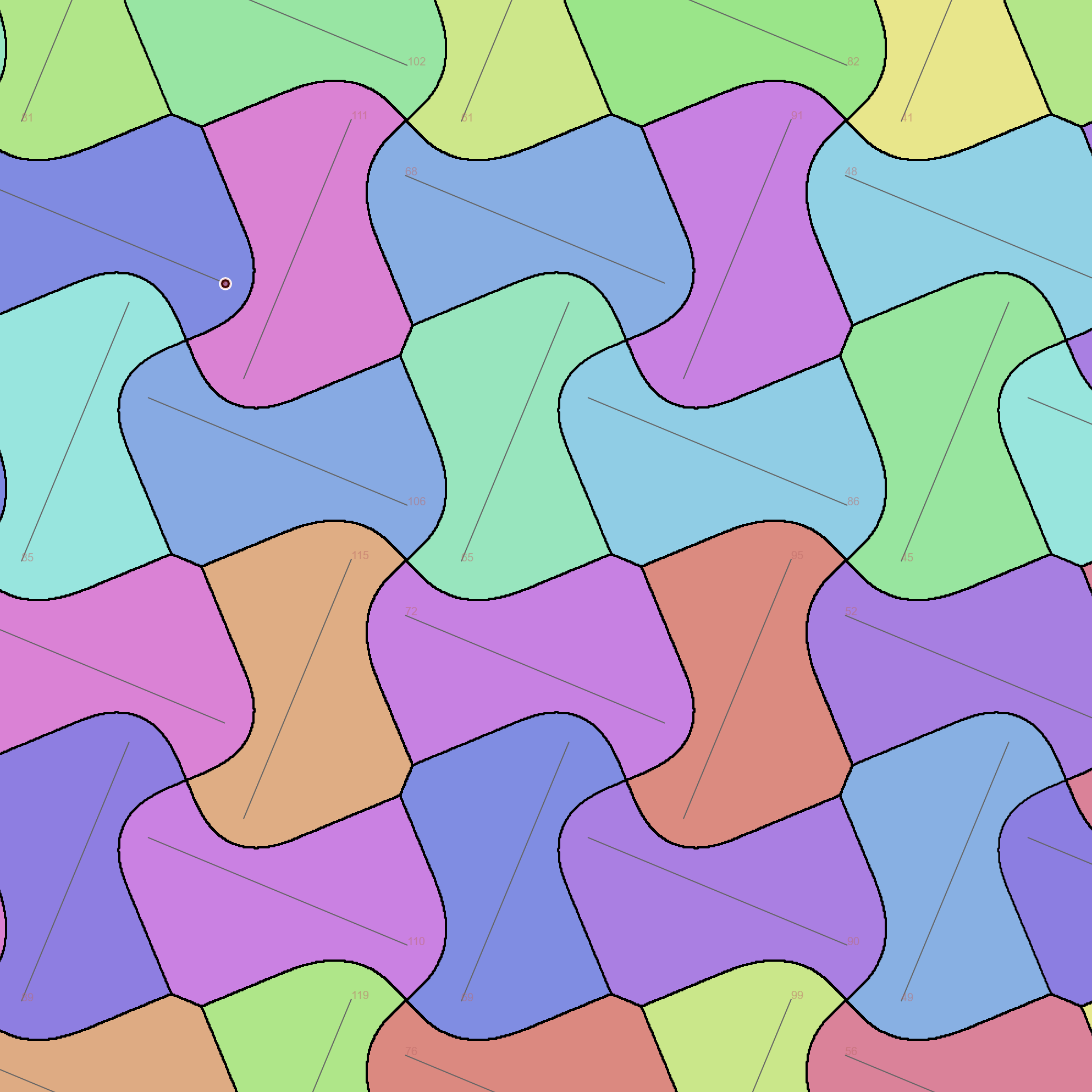}
        \caption{\it Patterns with  p4 symmetry.}
        \label{lines/p4_1}
    \end{subfigure}
\caption{\it Examples of curved tiles created by using single lines that are closed under p1, p2, pg, and p4 symmetries as Voronoi sites. The lines that are used as Voronoi sites are shown as light gray. These are the only ones of square-based symmetries (total 12)that can be used effectively to obtain curved boundaries with Voronoi decomposition. Other eight wallpaper symmetries, i.e. pm, cm, pmg, pgg, cmm, p4g, p4m, and pmm, include mirror operation and therefore, produce tiles with some straight boundary edges. In other words, these eight symmetries do not produce desired curved edges.}
\label{Square_Examples}
\end{figure}

\begin{figure}[htpb]
    \centering
    \begin{subfigure}[t]{0.48\textwidth}     \includegraphics[width=0.48\textwidth]{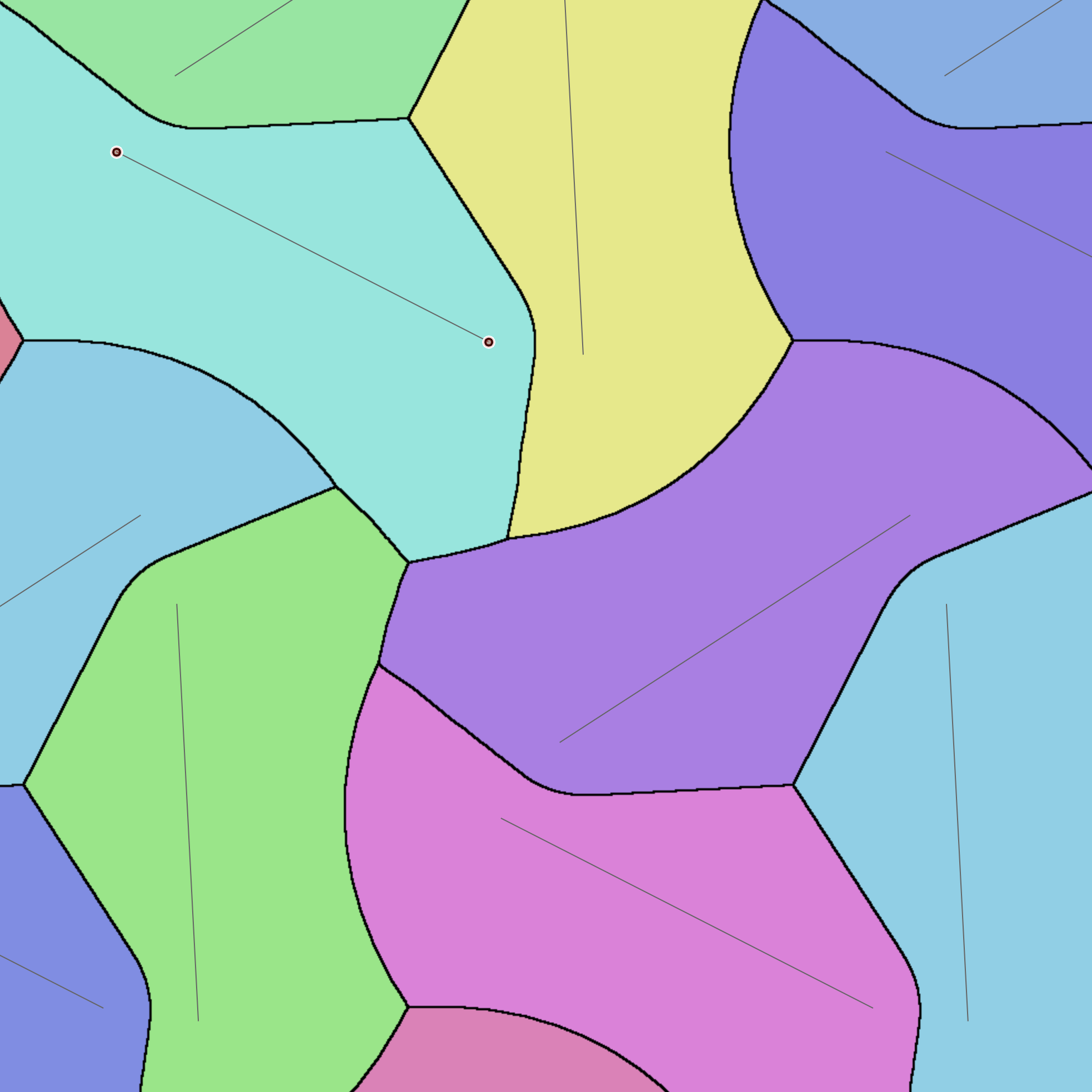}
        \includegraphics[width=0.48\textwidth]{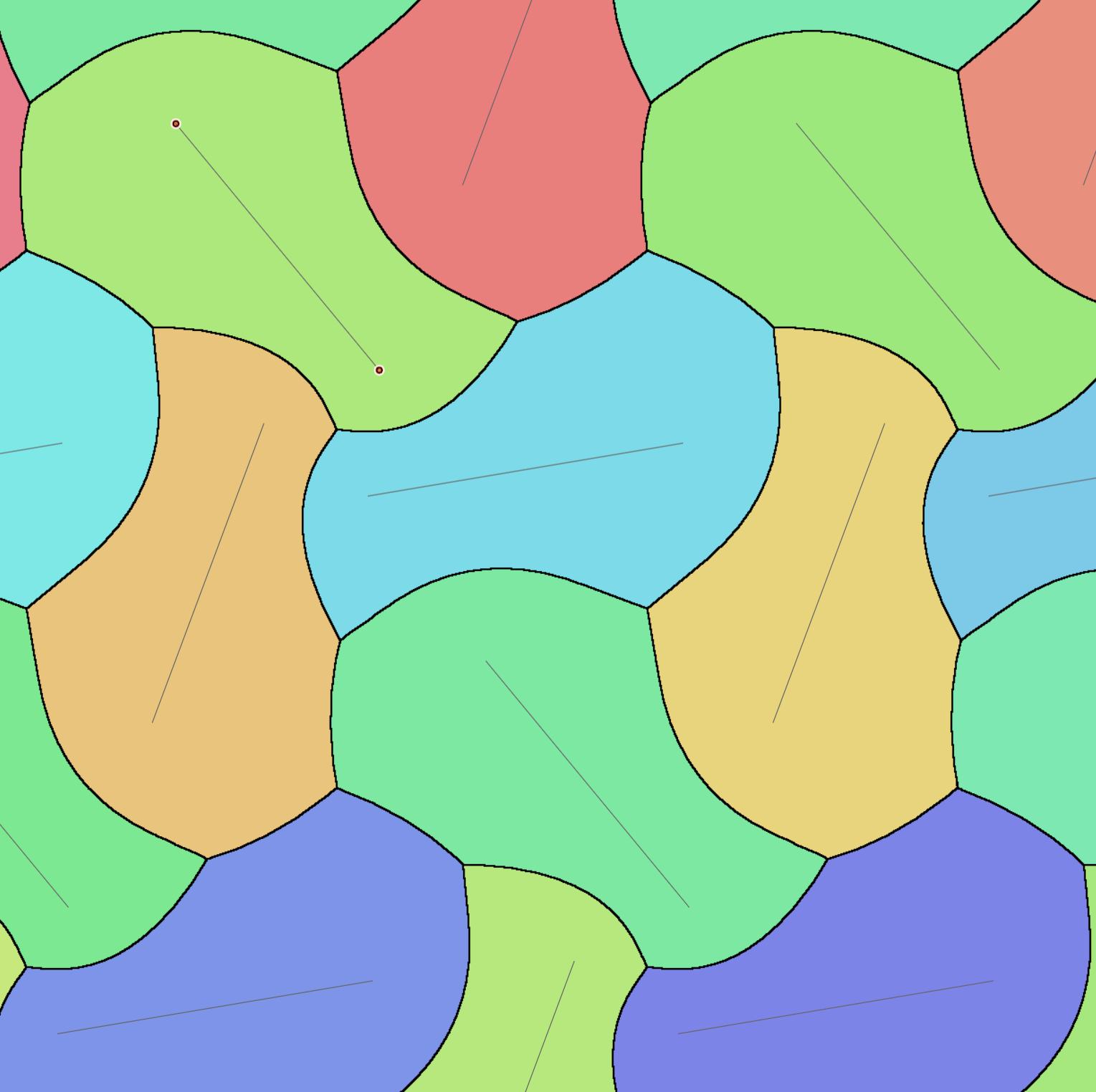}
        \caption{\it Patterns with p3 symmetry.}
        \label{lines/p3}
    \end{subfigure}
    \hfill
        \begin{subfigure}[t]{0.48\textwidth}
        \includegraphics[width=0.48\textwidth]{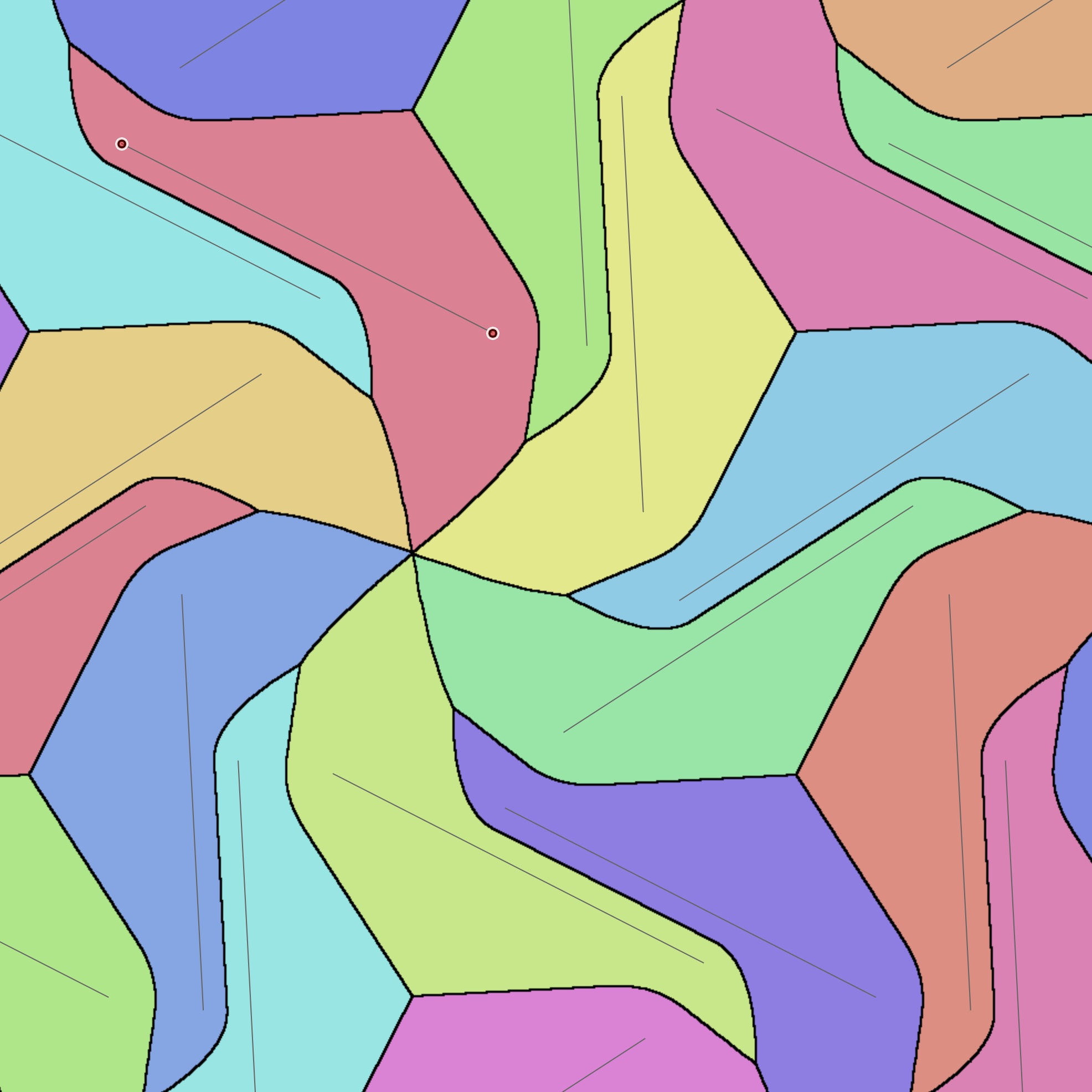}
        \includegraphics[width=0.48\textwidth]{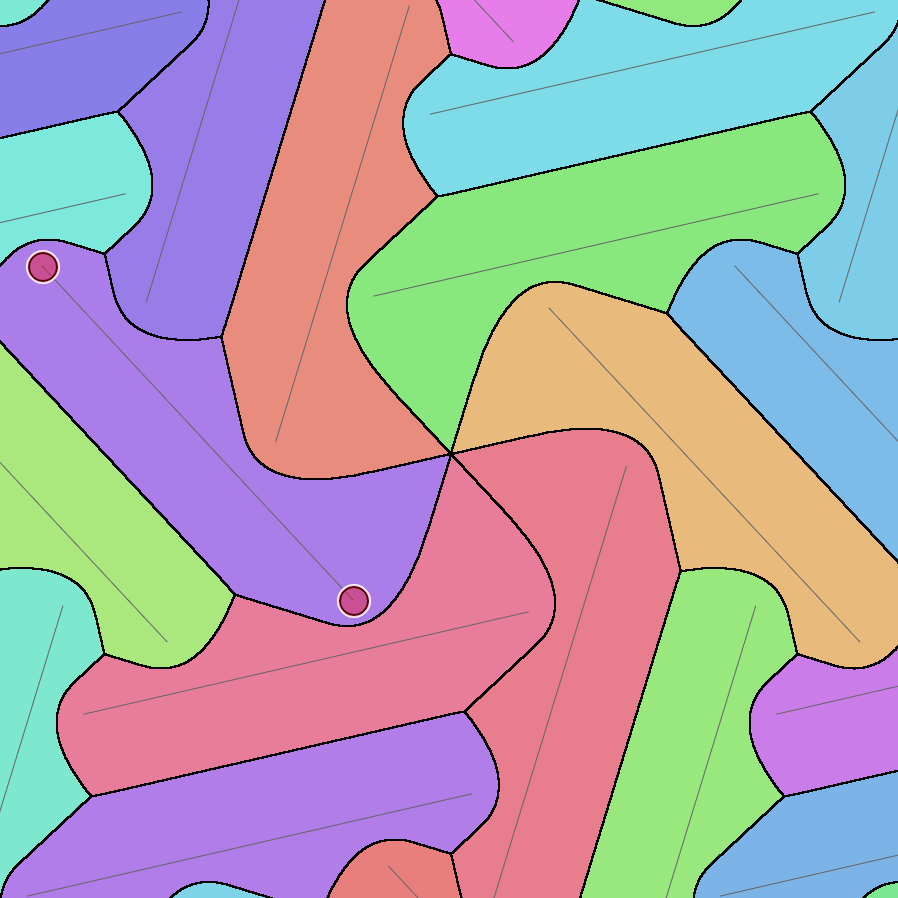}
        \caption{\it Patterns with p6 symmetry.}
        \label{lines/p6_0}
    \end{subfigure}
    \hfill
\caption{\it Examples of curved tiles created by using single lines that are closed under p3, and p6 symmetries as Voronoi sites. The lines that are used as Voronoi sites are shown as light gray. These are the only ones of hexagon-based symmetries (total 5) that can be used effectively to obtain curved boundaries with Voronoi decomposition. Other wallpaper symmetries,  p31m, p3m1, and p6m, include mirror operation and therefore, do not produce interesting tiling.}
\label{Hexagonal_Examples}
\end{figure}

\begin{figure}[!htpb]
    \centering
        \begin{subfigure}[t]{1.0\textwidth}
        \includegraphics[width=0.32\textwidth]{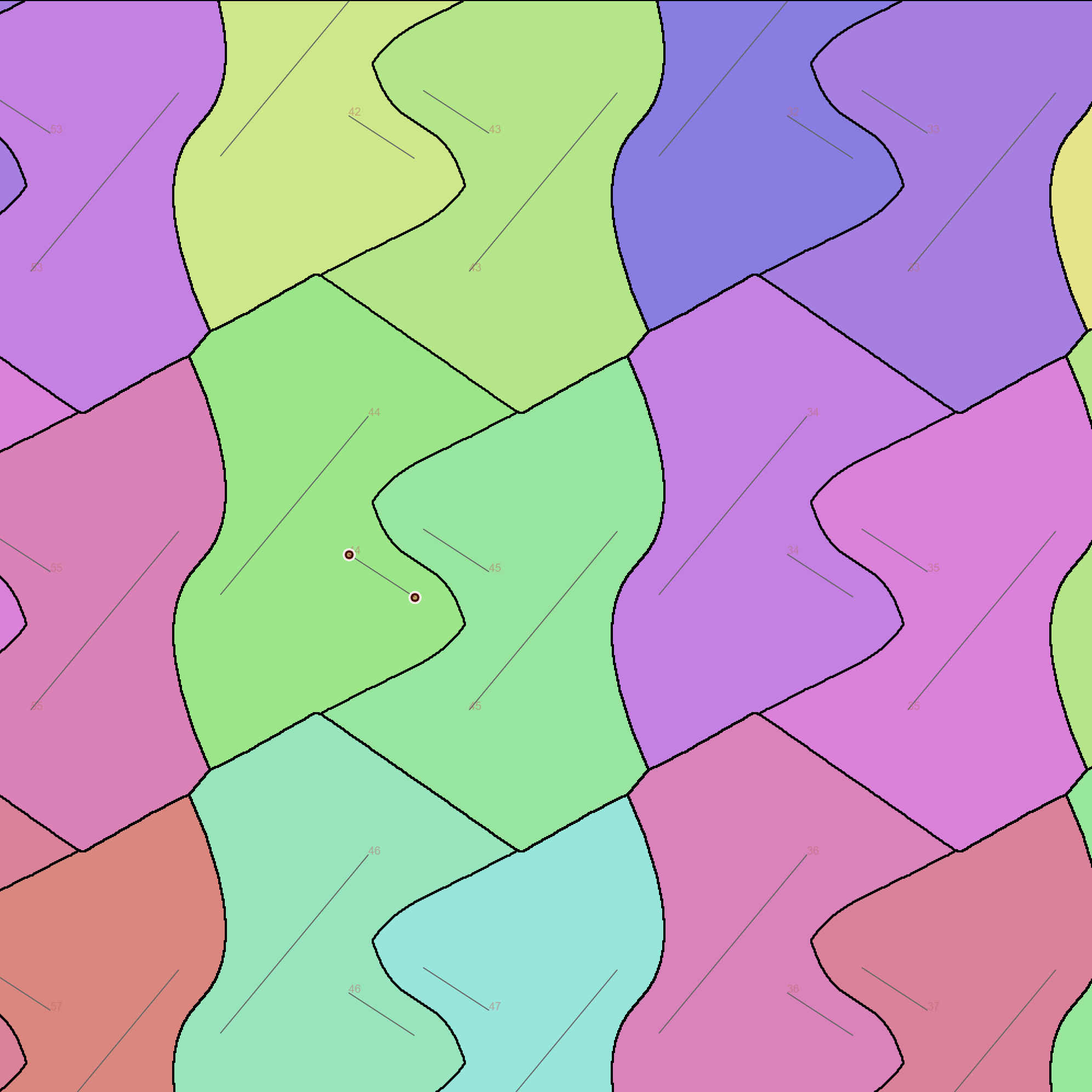}
        \includegraphics[width=0.32\textwidth]{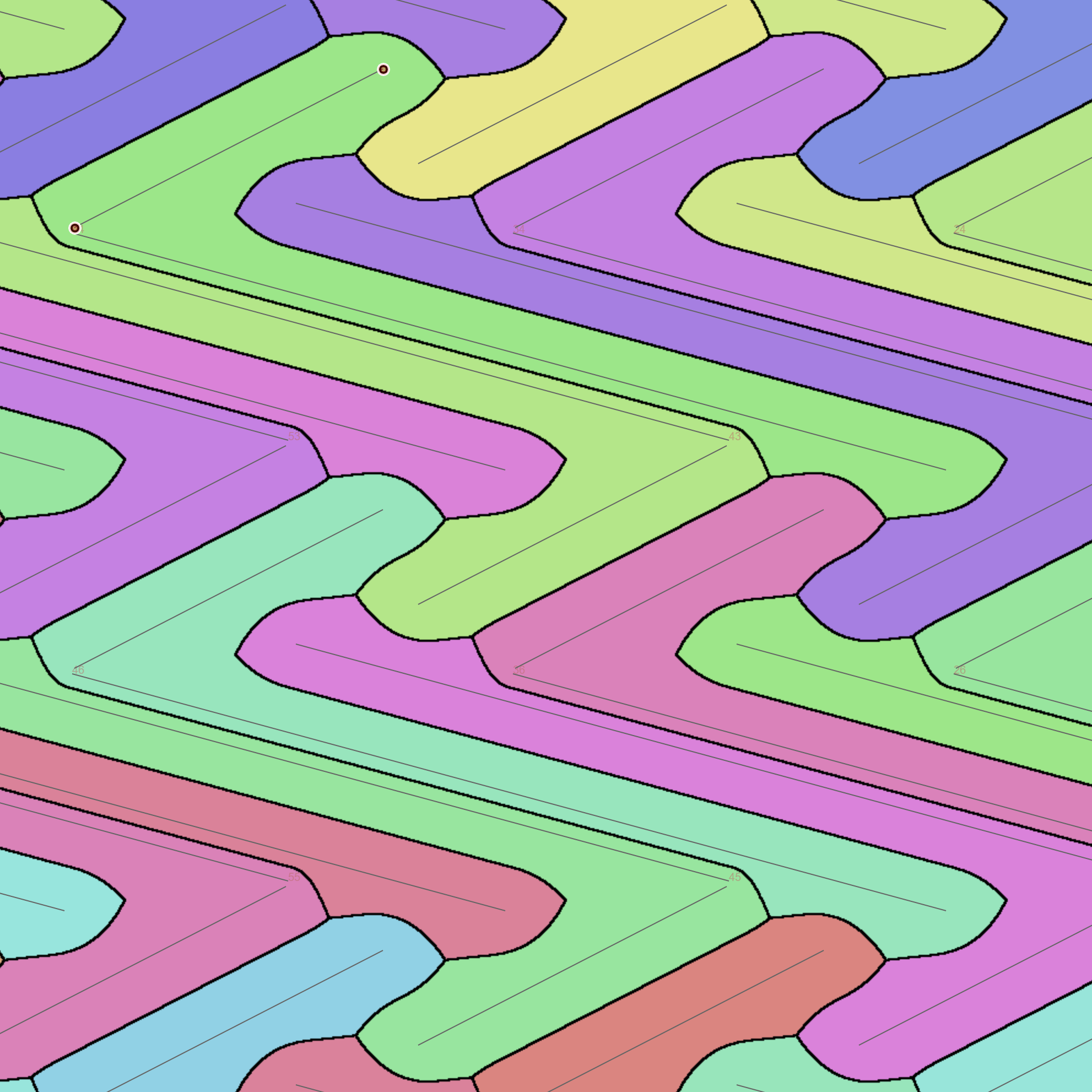}
        \includegraphics[width=0.32\textwidth]{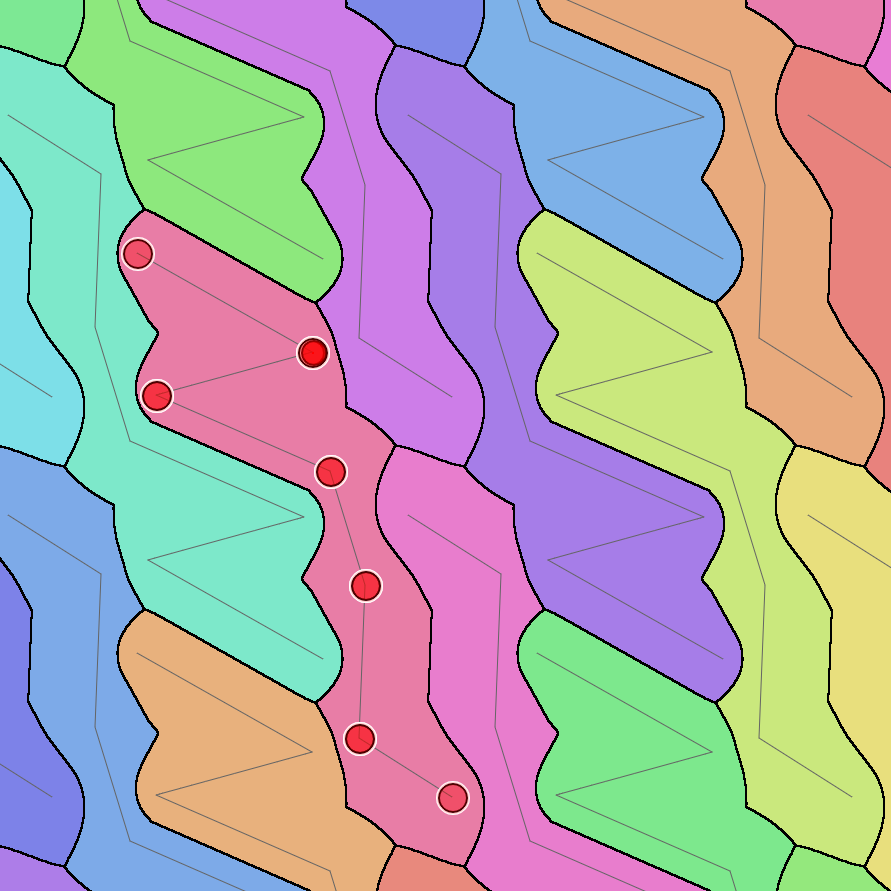}
        \caption{\it Patterns with p2 symmetry.}
        \label{multilines/p2}
    \end{subfigure}
    \hfill
    \begin{subfigure}[t]{0.48\textwidth}
        \includegraphics[width=1.0\textwidth]{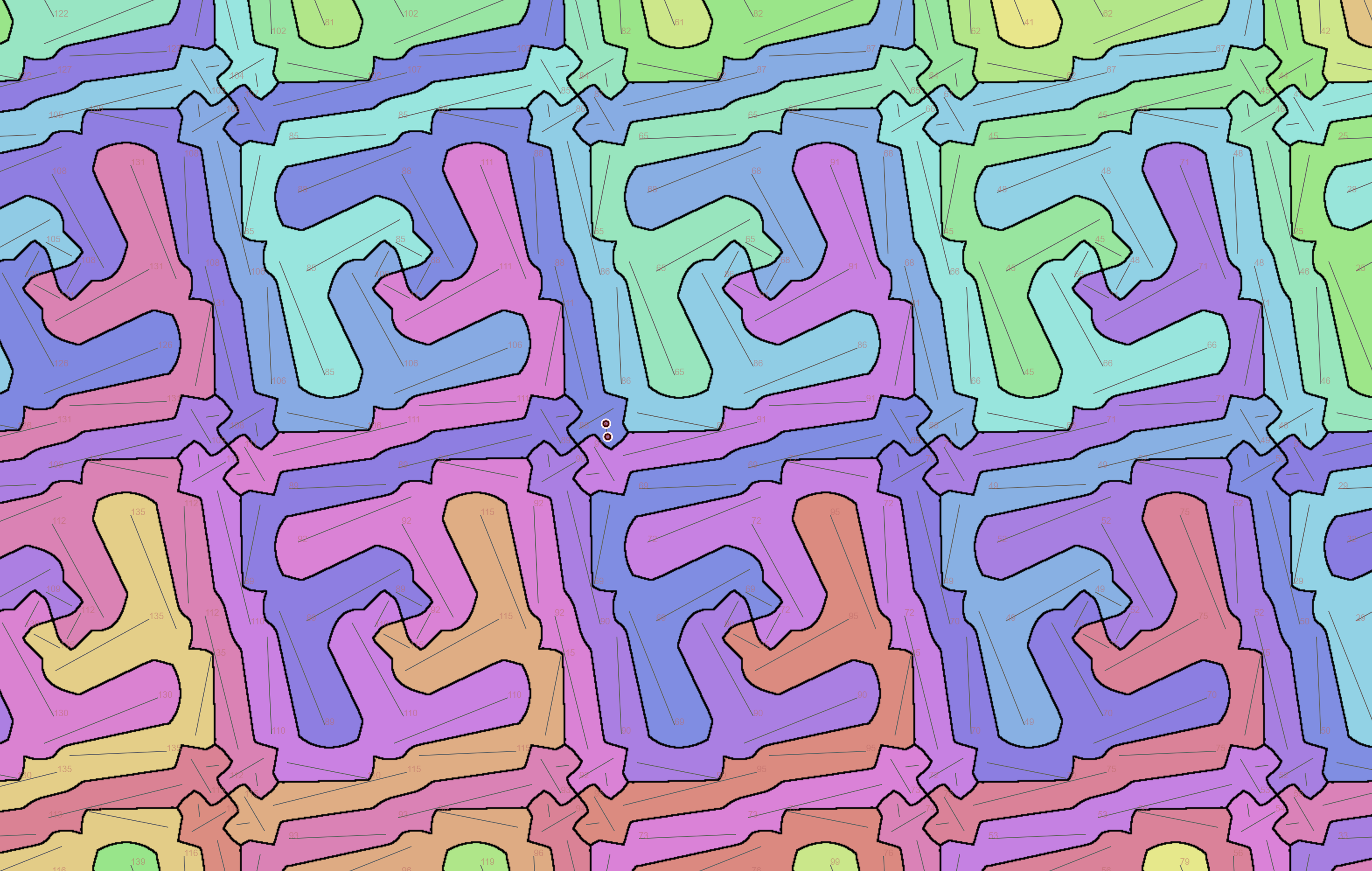}
        \caption{\it Pattern with p4 symmetry.}
        \label{multilines/p4_2}
    \end{subfigure}
    \hfill
        \begin{subfigure}[t]{0.48\textwidth}
        \includegraphics[width=1.0\textwidth]{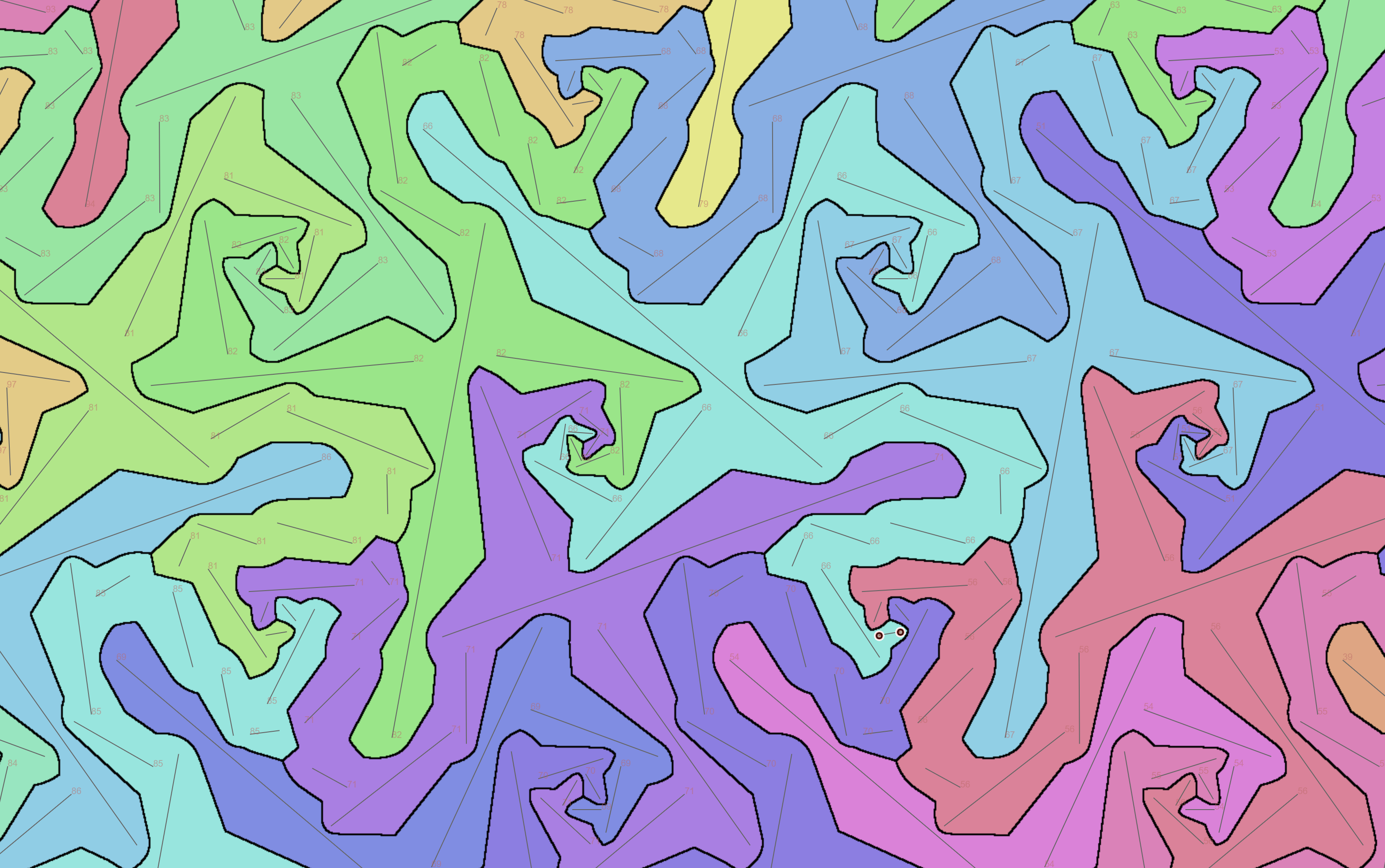}
        \caption{\it Pattern with p3 symmetry.}
        \label{multilines/p3_1}
    \end{subfigure}
\caption{\it More complicated examples using Voronoi sites that are a union of a set of lines. Using multiple lines more complex tiles can be obtained,  }
\label{Polyline_examples}
\end{figure}

\begin{figure}[!htpb]
    \centering
    \begin{subfigure}[t]{0.24\textwidth}
        \includegraphics[width=1.0\textwidth]{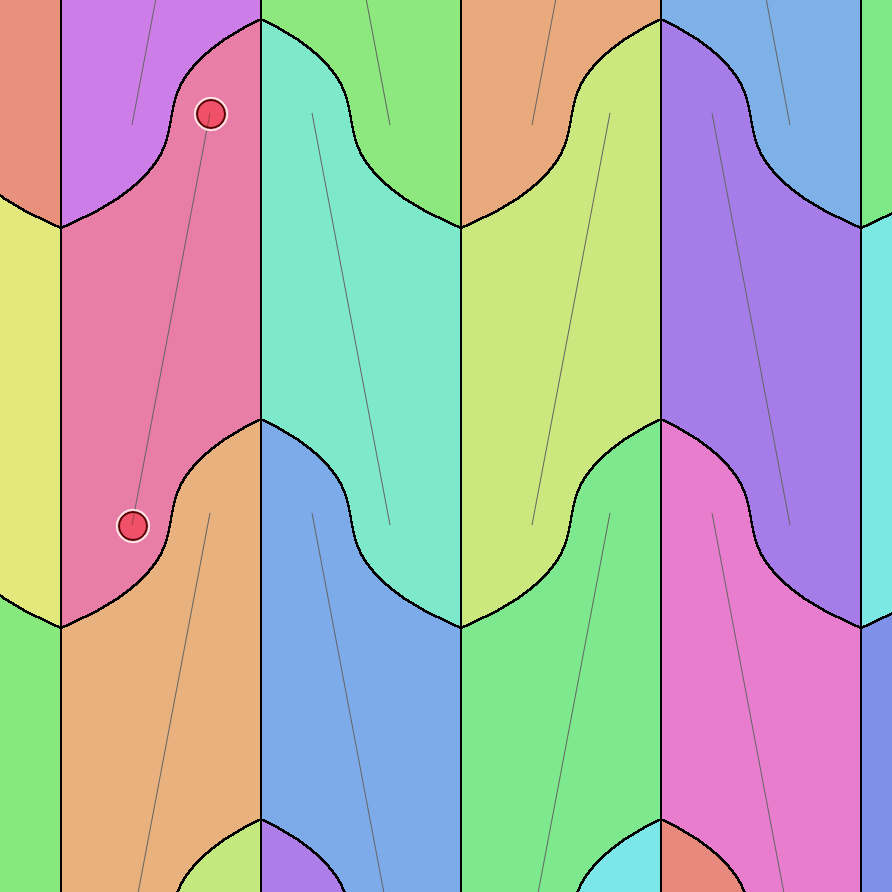}
        \caption{\it Example pattern with pm symmetry.}
        \label{failed/pm}
    \end{subfigure}
      \hfill
          \begin{subfigure}[t]{0.24\textwidth}
        \includegraphics[width=1.0\textwidth]{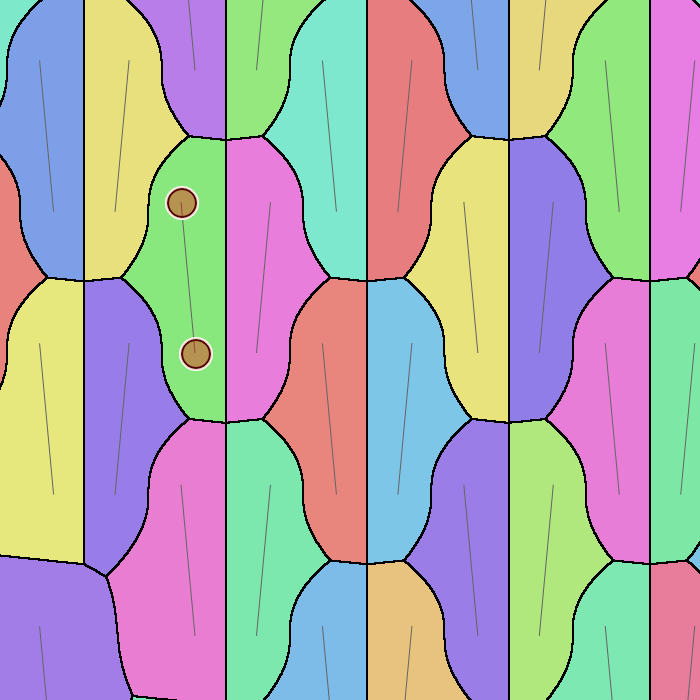}
        \caption{\it Example pattern with cm symmetry.}
        \label{failed/cm}
    \end{subfigure}
      \hfill
    \begin{subfigure}[t]{0.24\textwidth}
        \includegraphics[width=1.0\textwidth]{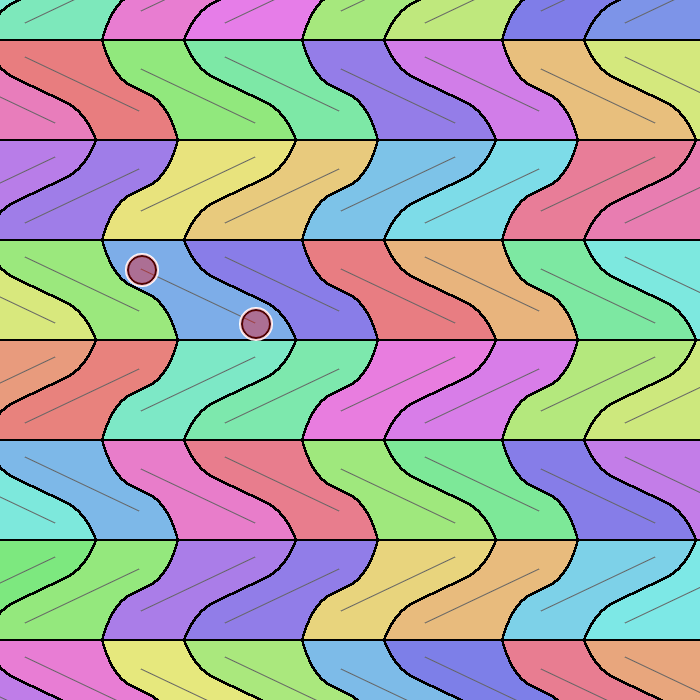}
        \caption{\it Example pattern with pmg symmetry.}
        \label{failed/pmg}
    \end{subfigure}
      \hfill
        \begin{subfigure}[t]{0.24\textwidth}
        \includegraphics[width=1.0\textwidth]{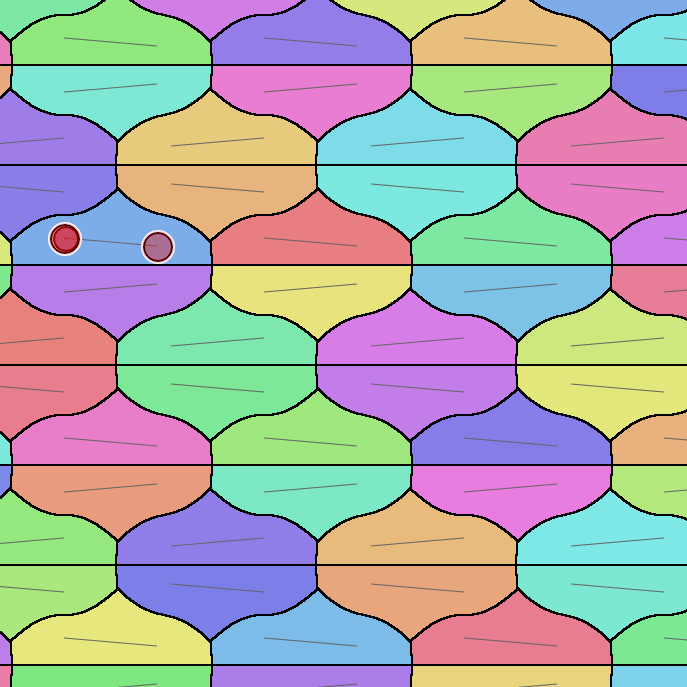}
        \caption{\it Example pattern with pgg symmetry.}
        \label{failed/pgg}
    \end{subfigure}
      \hfill
\caption{\it Examples that show the four symmetries that are not very useful since they cannot make all boundaries curved. Note that these four symmetries always produce in straight infinite lines regardless of how we choose Voronoi sites. }
\label{Failed_examples1}
\end{figure}

\begin{figure}[!htpb]
    \centering
    \begin{subfigure}[t]{0.32\textwidth}
        \includegraphics[width=1.0\textwidth]{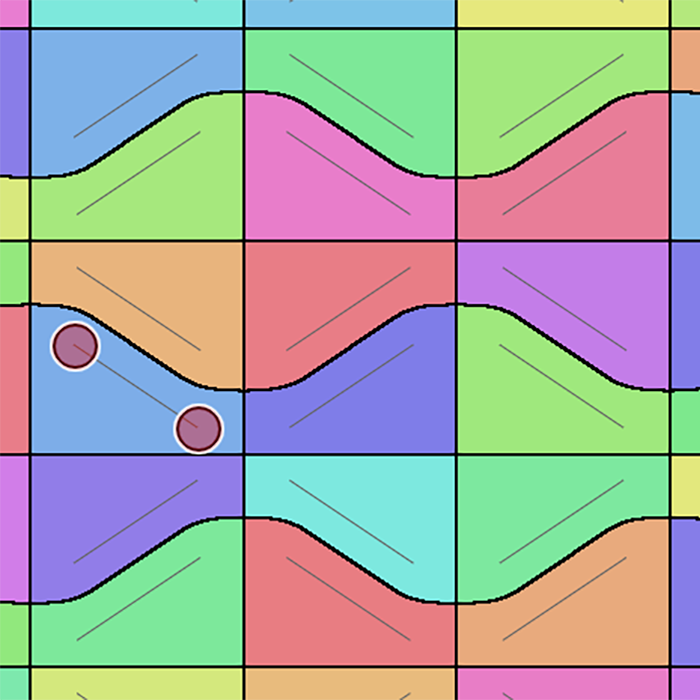}
        \caption{\it Example pattern with cmm symmetry.}
        \label{failed/cmm}
    \end{subfigure}
      \hfill
        \begin{subfigure}[t]{0.32\textwidth}
        \includegraphics[width=1.0\textwidth]{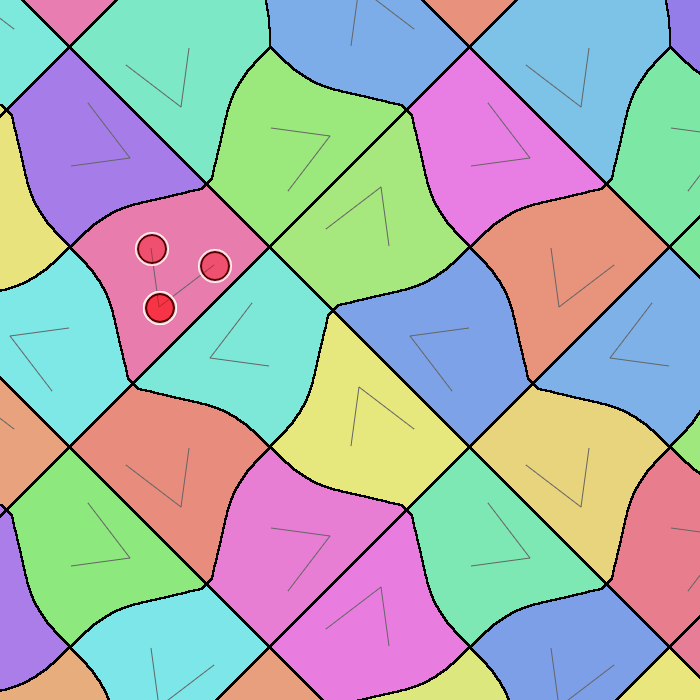}
        \caption{\it Example pattern with p4g symmetry.}
        \label{failed/p4g}
    \end{subfigure}
      \hfill
    \begin{subfigure}[t]{0.32\textwidth}
        \includegraphics[width=1.0\textwidth]{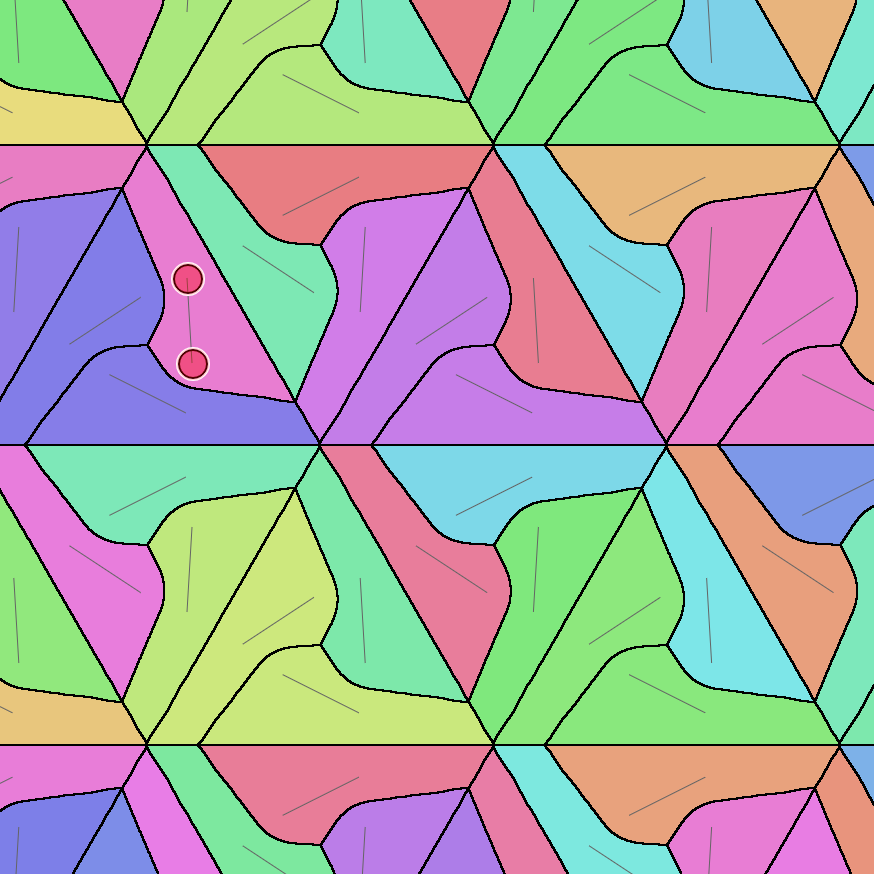}
        \caption{\it Example pattern with p31m symmetry.}
        \label{failed/p31m}
    \end{subfigure}
\caption{\it Examples that show the three symmetries that provide a decomposition of polygons. Two tiles in cmm create squares. Four tiles in p4g also create squares. Three tiles in p31 symmetry form equilateral triangles.  }
\label{Failed_examples2}
\end{figure}

\begin{figure}[!htpb]
    \centering
        \begin{subfigure}[t]{0.24\textwidth}
        \includegraphics[width=1.0\textwidth]{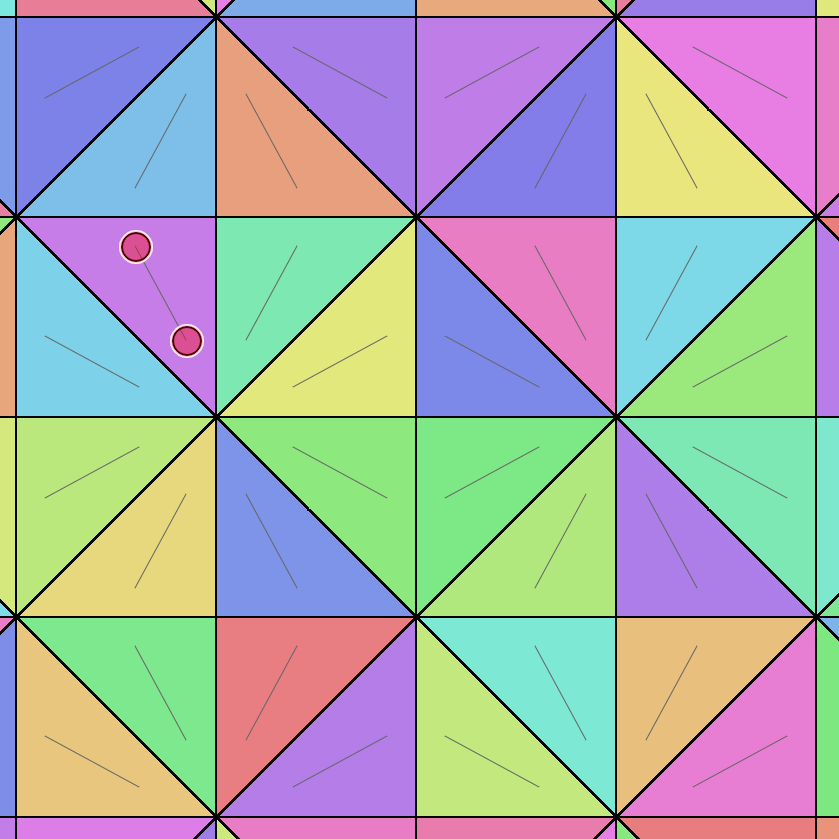}
        \caption{\it Example pattern with p4m symmetry.}
        \label{failed/p4m}
    \end{subfigure}
      \hfill
        \begin{subfigure}[t]{0.24\textwidth}
        \includegraphics[width=1.0\textwidth]{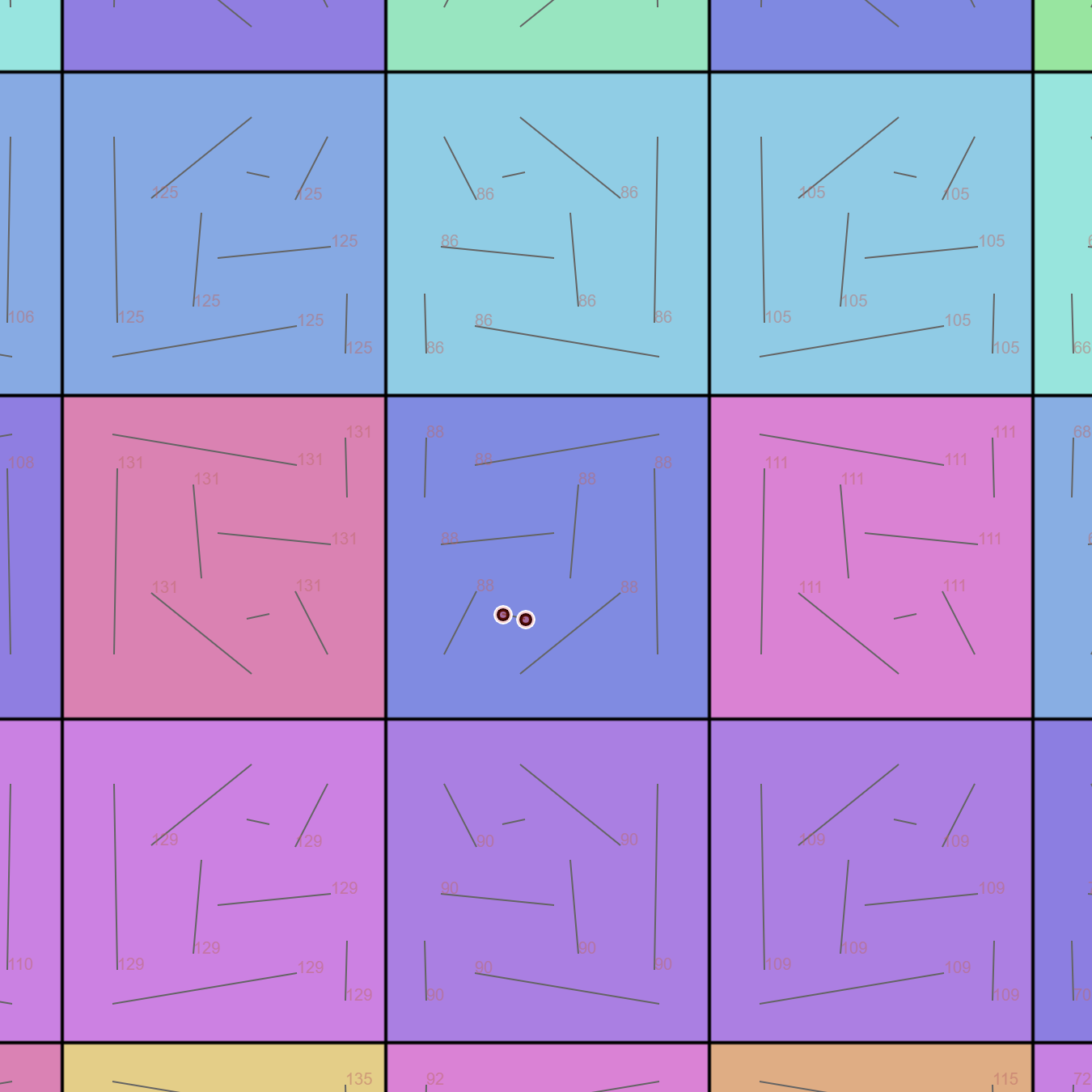}
        \caption{\it Example pattern with pmm symmetry.}
        \label{failed/pmm}
    \end{subfigure}
      \hfill
    \begin{subfigure}[t]{0.24\textwidth}
        \includegraphics[width=1.0\textwidth]{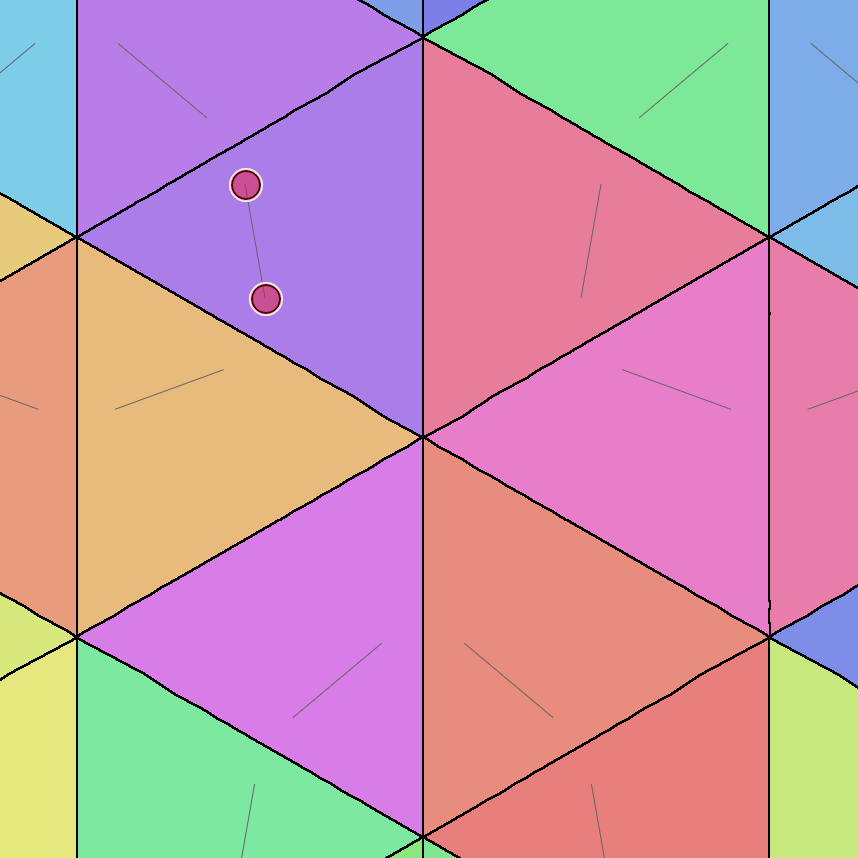}
        \caption{\it Example pattern with p13m symmetry.}
        \label{failed/p13m}
    \end{subfigure}
      \hfill
        \begin{subfigure}[t]{0.24\textwidth}
        \includegraphics[width=1.0\textwidth]{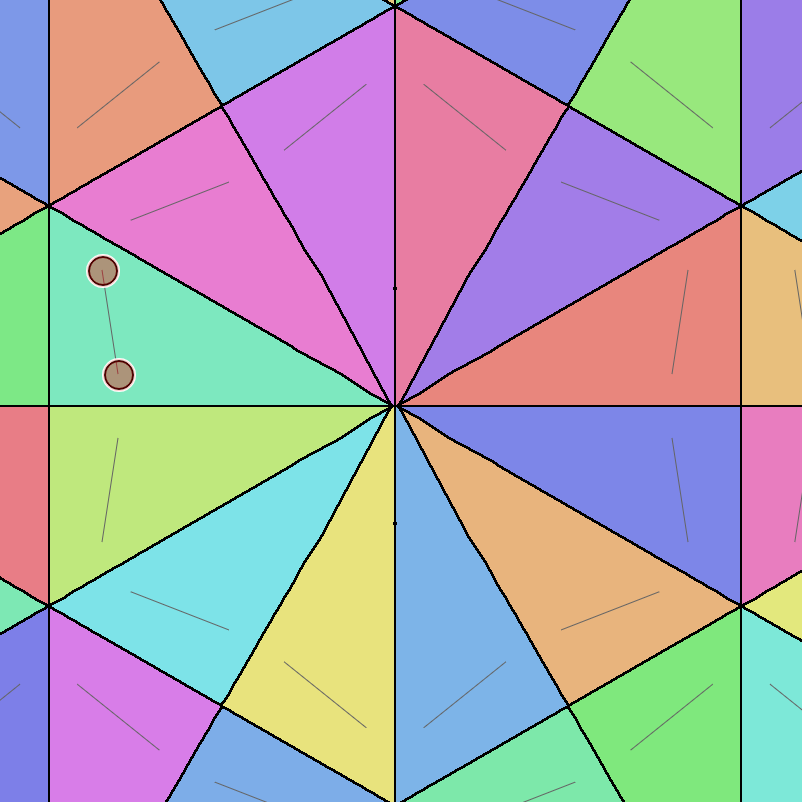}
        \caption{\it Example pattern with p6m symmetry.}
        \label{failed/p6m}
    \end{subfigure}
      \hfill
\caption{\it Examples that show four symmetries that always create the same polygon regardless of how many lines we use as Voronoi sites. The p4m symmetry only creates Right Isosceles Triangles. The pmm symmetry does not create anything beyond square packing. The p13m symmetry creates only Equilateral Triangles. The p6m symmetry creates right triangles that divide regular hexagons into 12 congruent shapes.  }
\label{Failed_examples3}
\end{figure}

The advantage of the second approach is that it can allow changes to the number of sides of polygonal tiles and allow the creation of multiple polygons \cite{amenta1995,akleman2000web}. 
Because of this flexibility, this approach appears to be the appropriate choice for designing symmetric tiles. However,  tiles are not just graphs, which are a set of arbitrary vertices, lines, and polygons. The tiles are cellular space-filling structures that come together without gaps and overlaps. If we do not guarantee the avoidance of gaps and overlaps, the user must be cautious not to include any gap or overlap when designing symmetric tiles. 

In this work, we present a new approach to developing flexible symmetric drawing systems that can guarantee no gaps and overlaps. Our approach is based on the Voronoi tessellations of 2D space by using the Voronoi sites closed on symmetry operations. This approach ensures that we obtain tiles that come together without gaps and overlaps or special attention from the user. This is really a generalization of Delaunay's original intention for the use of the Delaunay diagrams. Delaunay was the first to use symmetry operations on points and the Voronoi tessellations to produce space-filling polyhedra, which he called  Stereohedra \cite{delaunay1961,schmitt2016}. 

In this paper, we have extended his approach using higher-order Voronoi sites, i.e., lines, polylines, and curves, to construct space-filling tiles in 2D with curved boundaries (See Figures~\ref{Square_Examples}. and \ref{Hexagonal_Examples} for examples of using lines that are closed under some symmetry operations as Voronoi sites). Generally, 2D space is divided into regions so that each point $P$ in the region is closest to the Voronoi site $S$, which is also a point. Here, the distance is the length of the line that joins the point $P$ and the Voronoi site $S$. In our case, each Voronoi site is essentially a line segment $AB$, and space is tessellated using the closest distance of a point $P$ in the region to the line segment $AB$ instead of the conventional point. We assume the distance of a point $P$ to a line segment $AB$ is the length of the shortest line that can be drawn that connects the line segment $AB$ and the point $P$. We obtain Voronoi decompositions with polyline and curve Voronoi sites by using unions of line-based distance functions. 

\section{Related Work} 

A space-filling shape is a cellular structure whose replicas together can fill all of the space watertight, i.e., without having any voids between them \cite{loeb1991}. Equivalently, a space-filling shape is a cellular structure that can generate a tessellation of space \cite{grunbaum1980}. The 2D space-filling shapes have been widely used in Architecture since antiquity. They are frequently used by almost every civilization as wallpapers, wall decorations, ceilings, floor tiles, street pavements, and even facades
of buildings \cite{akleman2000web}. The symmetric patterns in Alhambra, Granada are probably the most well-known architectural usage of 2D space-filling shapes \cite{du2008symmetry}.

In this work, we developed an approach to provide a web-based space-filling symmetric tile design system. Two interesting cases of 2D tilings relevant to our approach are those presented by Kaplan~\cite{Kaplan2000} showing a wide variety of artistic patterns using specific Voronoi site configurations and Rao~\cite{rao2017} that show a systematic construction of 2D pentagonal tilings. In a sense, our work expands on these two works to move beyond polygonal tilings in 2D space to a rich design of curved symmetric tilings.

This work can also be a specialized version of recent works on the construction of 2.5D and 3D space-filling structures using higher-order Voronoi sites, namely Delaunay Lofts 
\cite{subramanian2019delaunay}, Generalized Abeille Tiles \cite{akleman2020generalized}, Bi-axial woven tiles \cite{krishnamurthy2020geometrically,yildiz2023modular}, and Voronoodles \cite{ebertvoronoodles,mullins2022voronoi}. modular These more powerful approaches that are designed for higher dimensional tiles \cite{krishnamurthy2023systems} are not reachable for most people. The systems require higher dimensional Voronoi decomposition and it is hard to turn them into an interactive web-based system. 
This particular work provides an interactive 2D tile design with a real-time interface. The system is available at \href{https://voronoi.viz.tamu.edu/}{https://voronoi.viz.tamu.edu/}.

The theoretical foundation for our curved tiles comes from the concept of symmetric Delone sets, attributed to Boris Delone (Delaunay). The original (non-symmetric) Delone sets essentially consist of a well-spaced set of points. Let $S$ denote a set of points in 2D Euclidean space, $\Re^{2}$. The $S$ is called a Delone set if it is both uniformly discrete and relatively dense \cite{delone1970new}. Formally, let $r_1 > r_0 > 0$ be two positive numbers. $S$ is uniformly discrete if each ball of radius $r_0$ contains at most one point in $S$. $S$ is relatively dense if every ball of radius $r_1$ contains at least one point of $S$ \cite{delone1976local}. If we use the points in $S \in \Re^2$ as Voronoi sites, one would obtain 2-Honeycombs that contain similar-sized convex polyhedra as Voronoi cells. Owing to this property, Delone sets, and related Meyer sets have been used to define tessellations \cite{lagarias1999geometricI, lagarias1999geometricII, lagarias2003repetitive, lagarias1996meyer}. 

The symmetric Delone sets in 2D are invariant with respect to one of the 17 wallpaper groups \cite{dolbilin2015delone}. Therefore, an ideal space filling 2D tiles can be described by the Delone sets in 2D Euclidean space along with a wallpaper group of Euclidean isometries acting on this point  \cite{delaunay1961}.
A Delone set $S$ is symmetric in 2D if, for every two points $\mathbf{p}, \mathbf{q} \in S$, there exists a rigid motion of 2D space that takes  $S$ to  $S$ and $\mathbf{p}$ to $\mathbf{q}$. The standard mathematical model of an ideal tesselation also involves symmetric Delone sets \cite{dolbilin2015delone,dolbilin2021regularity}. The congruent polygonal tiles with straight edges are guaranteed to be obtained by the Voronoi decomposition of points in a symmetric Delone set.  
The underlying principle for tesselations is that if a 2D arrangement of given a set of Voronoi sites is symmetric Delone, then the Voronoi tessellation results in a unique repeatable space-filling polygon with straight edges. 

\section{Delone sets with 1-manifolds  with boundaries in 2D} 

Our key observation in this paper the Voronoi sites in symmetric Delone sets need not be points. They can be 1-manifolds or 1-manifolds  with boundaries in 2D space such as line segments, open or closed polylines, and open and closed curves. In this paper, we only experiment with 1-manifolds with boundaries, in other words, line segments, polylines, and curve segments. We also assume that these polylines and curve segments are simple. In other words, these shapes do not self-intersect with themselves. 

We now need to generalize the concept of the original Delone sets with points to Delone sets with such ``simple'' 1-manifolds with boundaries. For Delone sets, we essentially need a well-spaced set of 1-manifolds with boundaries filling 2D space. Now, let $\mathbf{C}$ denote a set of 1-manifolds with boundaries with exactly the same geometric shape in 2D Euclidean space, $\Re^{2}$. Note that for such simple 1-manifolds with boundaries, the distance to the shape is well-defined. Let $\mathbf{c} \in \mathbf{C}$ denote a simple 1-manifold with boundary and $p \in \Re^2$, then let the function $D(p,\mathbf{c}$ denote the distance of point $p$ to the $\mathbf{c}$. Now, assume we define a generalized ball of radius $r$ as follows: 
$$B = \{ p=(x,y) | D(p,\mathbf{c} \leq r.$$
For lines, this ball will have the shape of a capsule as shown in Figure~\ref{renders/capsule}. Piecewise linear 1-manifolds that consist of multiple lines will produce a union of these capsules as generalized balls. 
Using these generalized balls, we can redefine the terms uniformly discrete and relatively dense for Delone sets with such ``simple'' 1-manifolds with boundaries. Similarly, 
Let $r_1 > r_0 > 0$ be two positive numbers. 
$S$ is uniformly discrete if there exist an $r_0$ such that each ball of radius $r_0$ contains at most shape in $\mathbf{C}$. Similarly, $\mathbf{C}$ is relatively dense if every ball of radius $r_1$ contains at least one shape of $\mathbf{C}$. We call $\mathbf{C}$ a Delone set if it is both uniformly discrete and relatively dense. These non-symmetric Delone sets essentially consist of a well-spaced set of the same type of 1-manifolds with boundaries.

We also need to define symmetric Delone sets for `simple'' 1-manifolds with boundaries. For symmetry, the Delone set $\mathbf{C}$ must be invariant with respect to one of the 17 wallpaper groups. Formally, 
the Delone set $\mathbf{C}$ is symmetric if, for every two points $\mathbf{c_0}, \mathbf{c_1} \in \mathbf{C}$, there exists a rigid motion of 2D space that takes  $\mathbf{C}$ to  $\mathbf{C}$ and $\mathbf{c_0}$ to $\mathbf{c_1}$. We can then obtain congruent polygonal tiles with curved edges by the Voronoi decomposition of shapes in a symmetric Delone set $\mathbf{C}$.  In other words, the Voronoi tessellations of $\mathbf{C}$ can result in unique repeatable space-filling polygons with curved edges. 

With this in view, the main objectives of our framework are to (1) produce an arrangement of higher dimensional Voronoi sites based on a given wallpaper symmetry group, and (2) satisfy the conditions such that the arrangement will be symmetric Delone and (3) partition the space using Voronoi decomposition.

Note that, we focus only on non-mirror symmetry groups based on isometries of a cube. The main reason for this restriction is that operations that create chiral shapes such as mirrors are not practically useful. In multi-mirror operations, such as \textit{pmm} and \textit{cmm} in 2D and \textit{pmmm} and \textit{cmmm} in 3D, there is no difference between using high-dimensional shapes and a single point as Voronoi sites. Even if a mirror operation is used just once, we end up with planar (i.e. not-curved) interfaces regardless of the shape of Voronoi sites. We develop a framework to obtain all elements of symmetries induced by cube isometries. 

\subsection{Line Segments as Voronoi Sites}

\begin{figure}[!htpb]
    \centering
        \begin{subfigure}[t]{0.27\textwidth}
        \fbox{\includegraphics[width=1.0\textwidth]{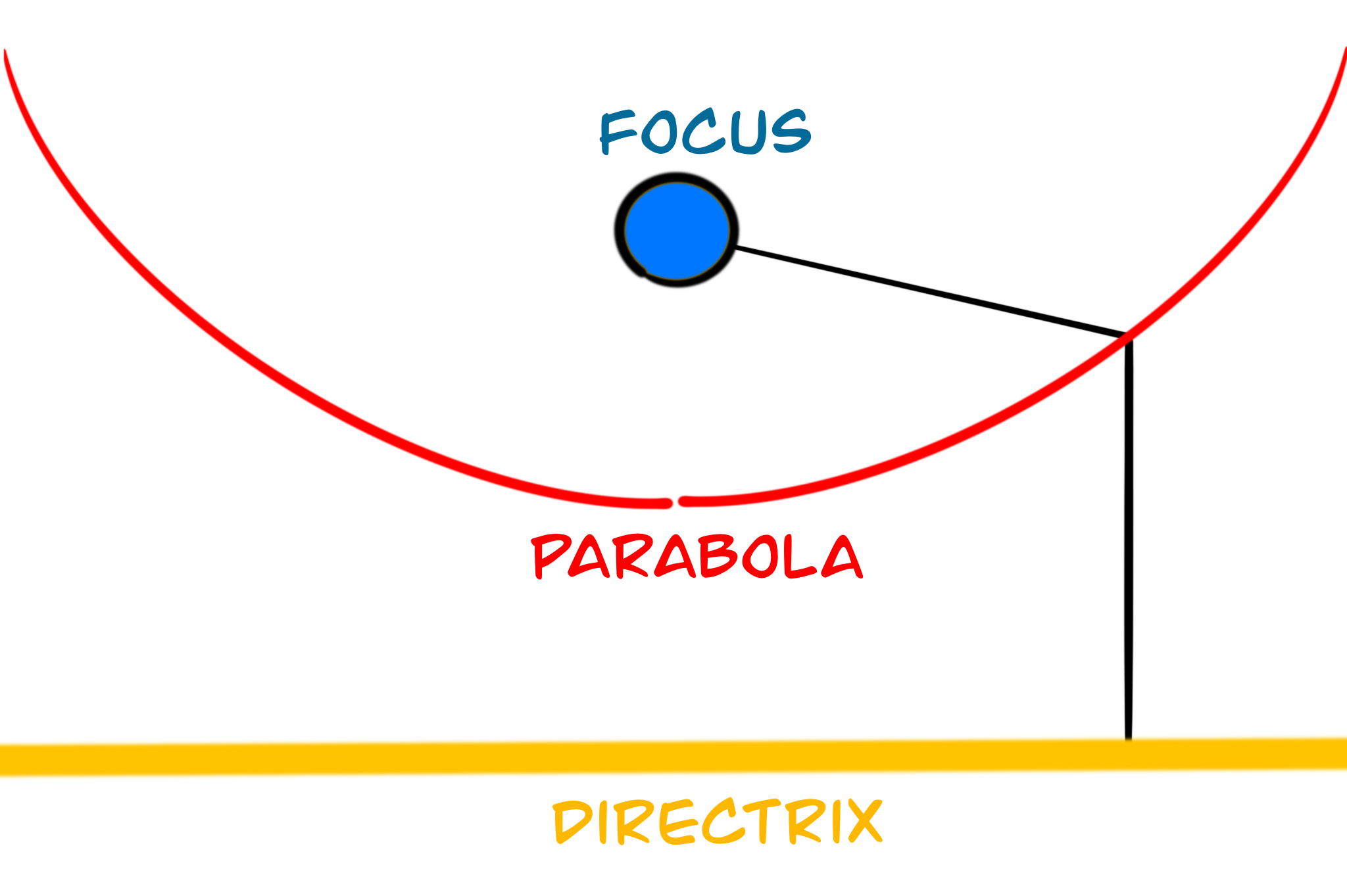}}
        \caption{\it Infinite Line as Directix produces a complete parabola.}
        \label{renders/InfiniteLine}
    \end{subfigure}
      \hfill
        \begin{subfigure}[t]{0.31\textwidth}
        \fbox{\includegraphics[width=1.0\textwidth]{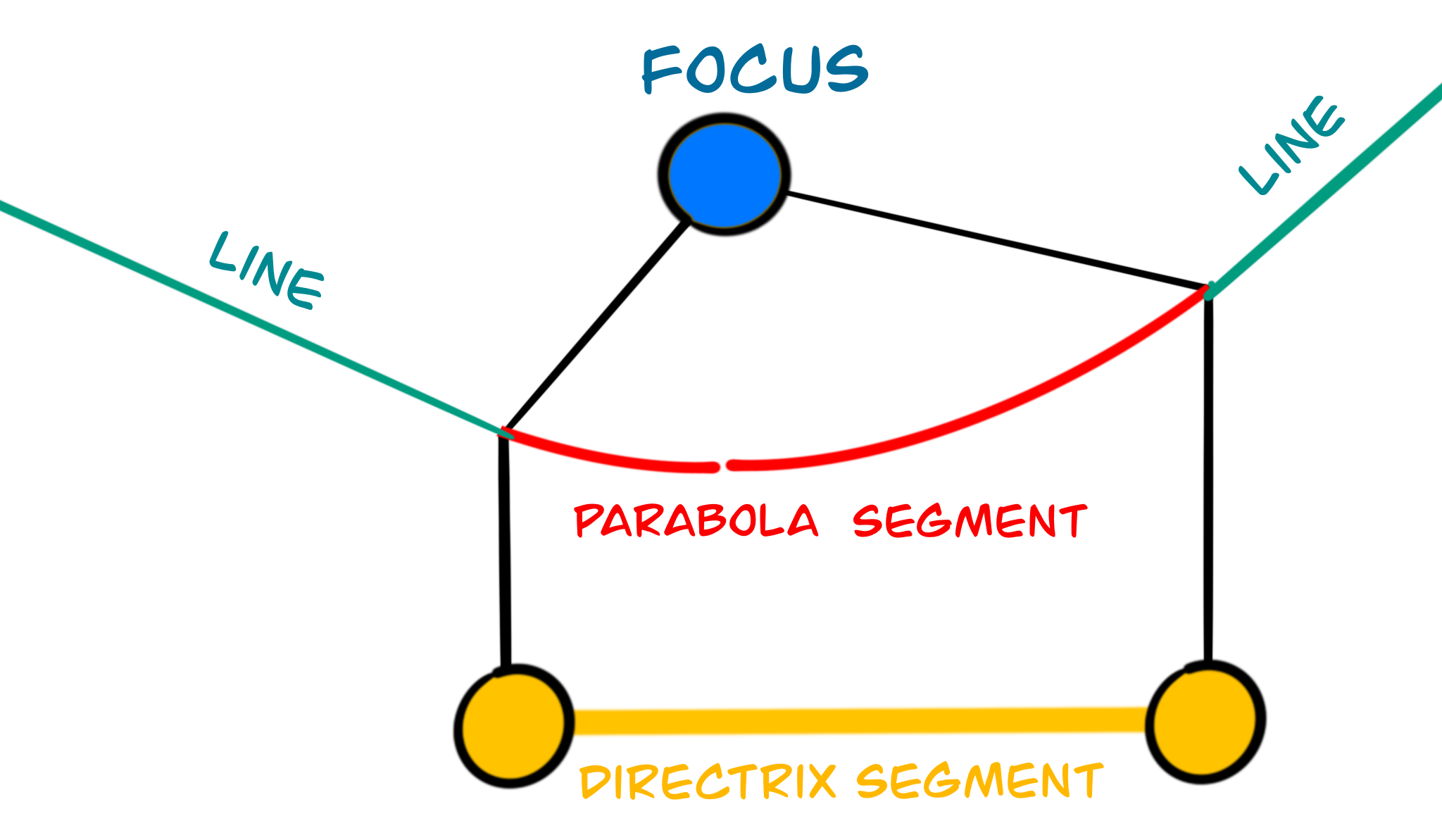}}
        \caption{\it A line segment as Directix produces a piecewise parabolic curve.}
        \label{renders/LineSegment}
    \end{subfigure}
      \hfill
    \begin{subfigure}[t]{0.18\textwidth}
        \fbox{\includegraphics[width=1.0\textwidth]{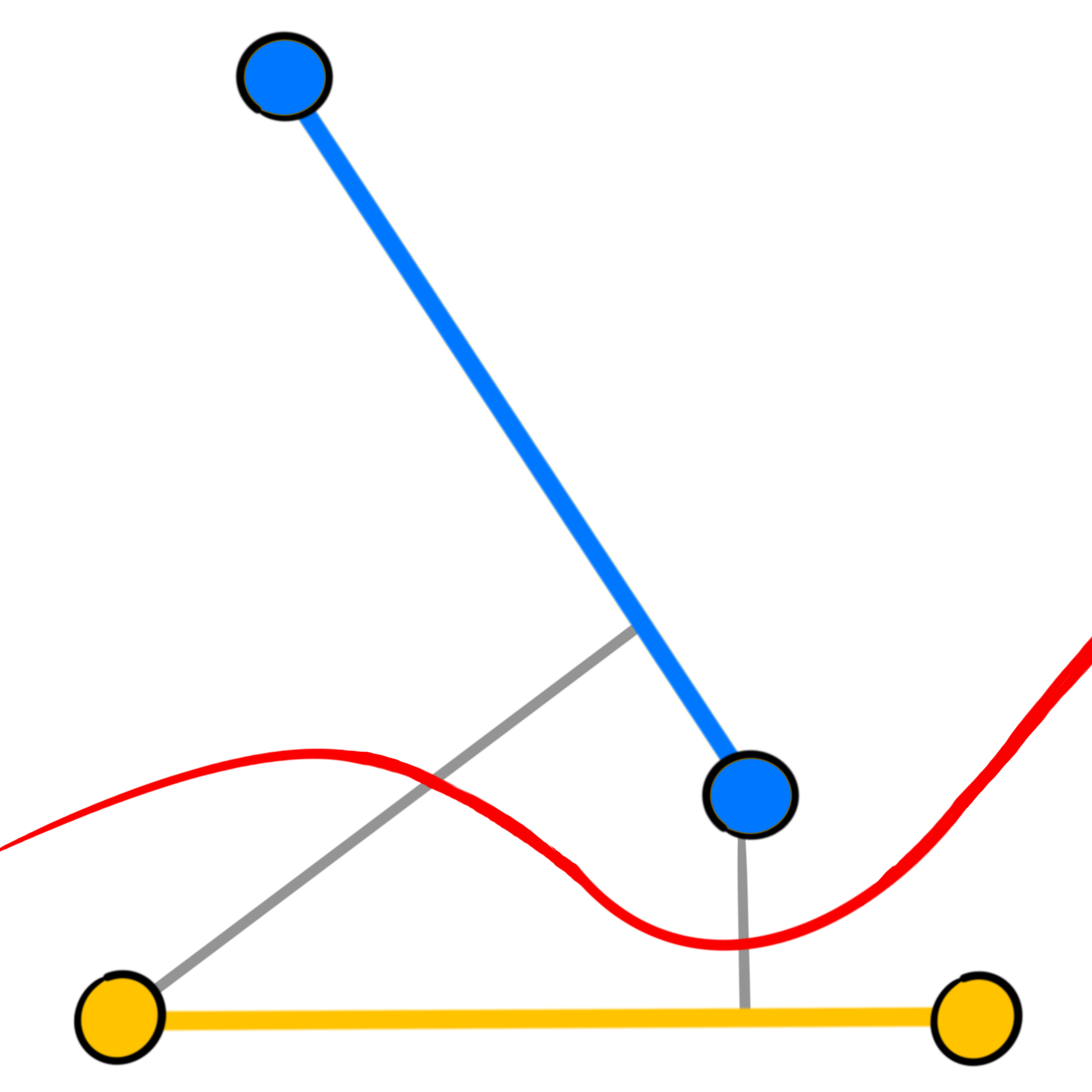}}
        \caption{\it Two random line segments.}
        \label{renders/TwoLineSegments}
    \end{subfigure}
      \hfill
    \begin{subfigure}[t]{0.18\textwidth}
       \fbox{\includegraphics[width=1.0\textwidth]{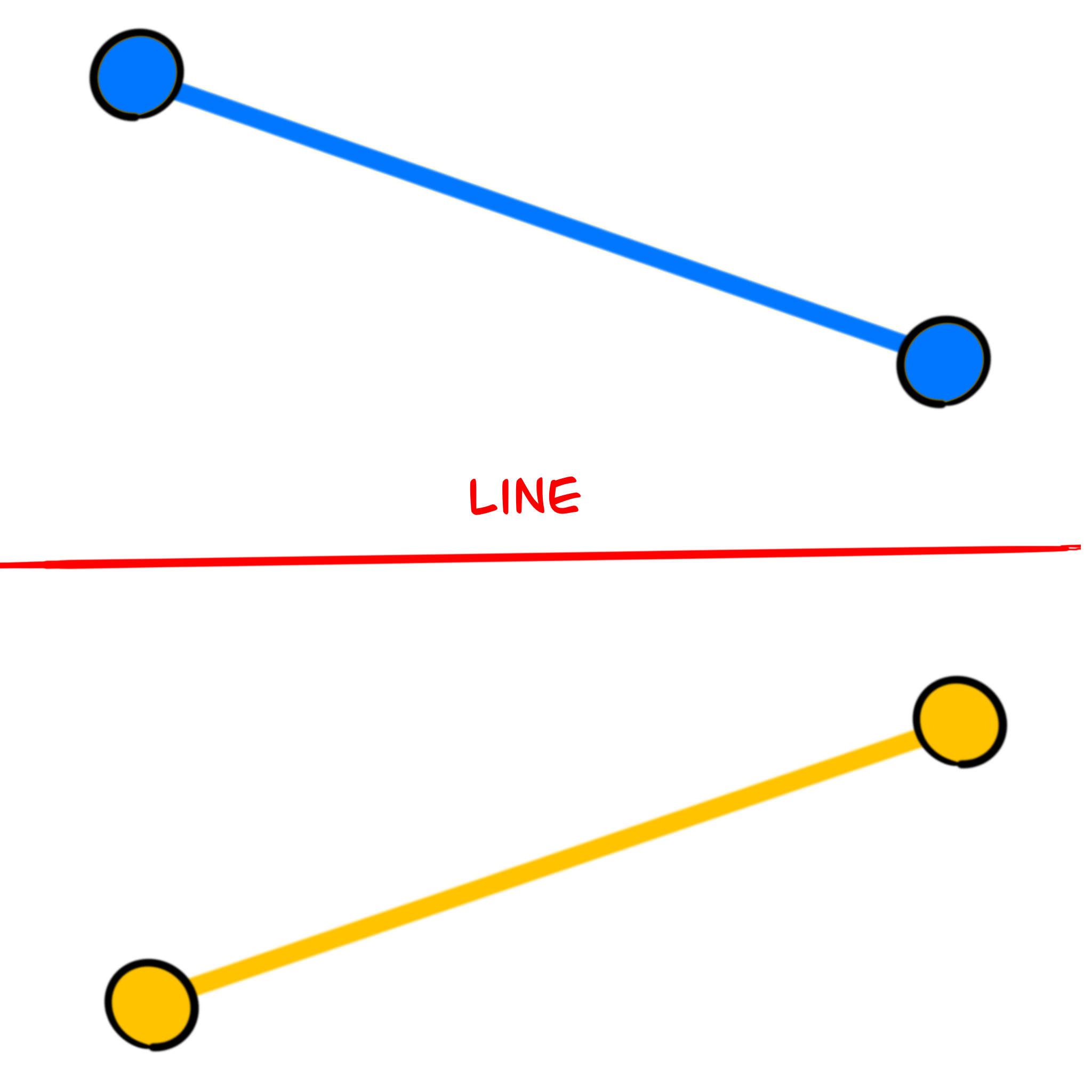}}
        \caption{\it Two mirrored line segments.}
        \label{renders/Mirror}
    \end{subfigure}
      \hfill
\caption{\it These figures show how parabolic Voronoi boundaries are obtained when using line segments as Voronoi sites. Note that a line segment as Directix produces a piecewise curve that consists of a parabolic segment along with two lines. Multiple line segments can produce more complex boundary that consists of multiple parabolic curves. On the other hand, if two line segments are mirrored, they only produce the original mirror line as the Voronoi boundary. }
\label{ParabolaGeneration}
\end{figure}

Let a line segment be represented by two endpoints $\textbf{p}_0$ and $\textbf{p}_1$ as follows
$$\textbf{p} = \textbf{p}_0 (1-t) + \textbf{p}_0 t $$
where $t \in [0,1]$. All lines in a symmetric Delone set can be obtained by transforming the original two points $\textbf{p}_0$ \& $\textbf{p}_1$, and creating a line between them. 
When all the Voronoi sites are line segments, the boundaries can be either paraboloid curves or lines. Note that the parabola is already defined as the locus (or collection) of points equidistant from a given point (the focus) and a given infinite line (the directrix) \cite{jones2023} (See Figure~\ref{renders/InfiniteLine}). If the directrix line is not infinite, i.e. a line segment, Voronoi decomposition produces only a segment of a parabola (See Figure~\ref{renders/LineSegment}). Voronoi decomposition with randomly oriented multiple line segments can produce even more complex boundary that consists of multiple parabolic curves (see Figure~\ref{renders/TwoLineSegments}) \cite{subramanian2019delaunay}. On the other hand, two mirrored line segments can only produce the original mirror line as Voronoi boundary as shown in
Figure~\ref{renders/Mirror}. 

Mirrored shapes (not only line segments) are a major issue to obtain with curved boundaries. If the symmetry operations include a mirror, we cannot avoid getting straight boundaries. Figures~\ref{Failed_examples1}, ~\ref{Failed_examples2}, and~\ref{Failed_examples3} show three types of symmetry operations that can produce straight lines. This problem is something that is not critical for Delone set-based methods since original Delone sets use only points to obtain Voronoi decomposition. The results, therefore, will always be a straight line. It is also important to note that p4m, pmm, p13m, and p6m are the worst cases since they can produce the same boundaries regardless of the shapes of Voronoi sites in Figure~\ref{Failed_examples3}. These four cases are also undesirable for the original point-based Delone set methods since they can not provide any control to creating tiles other than four polygons shown in Figure~\ref{Failed_examples3}. 

\section{Image Based Voronoi Tessellation with Lines as Voronoi Sites}

\begin{figure}[!htpb]
    \centering
    \begin{subfigure}[t]{0.19\textwidth}
        \includegraphics[width=1.0\textwidth]{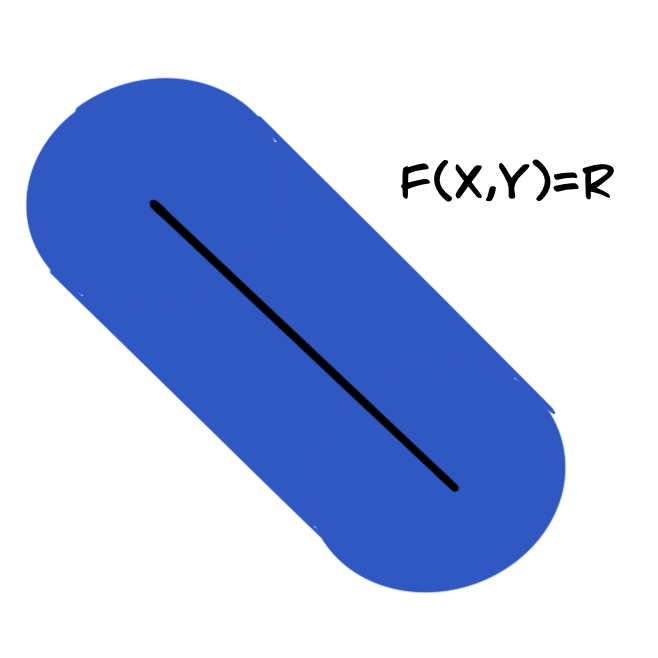}
        \caption{\it The shape of a ball that is defined as equidistant to a line.}
        \label{renders/capsule}
    \end{subfigure}
    \hfill
    \begin{subfigure}[t]{0.19\textwidth}
        \includegraphics[width=1.0\textwidth]{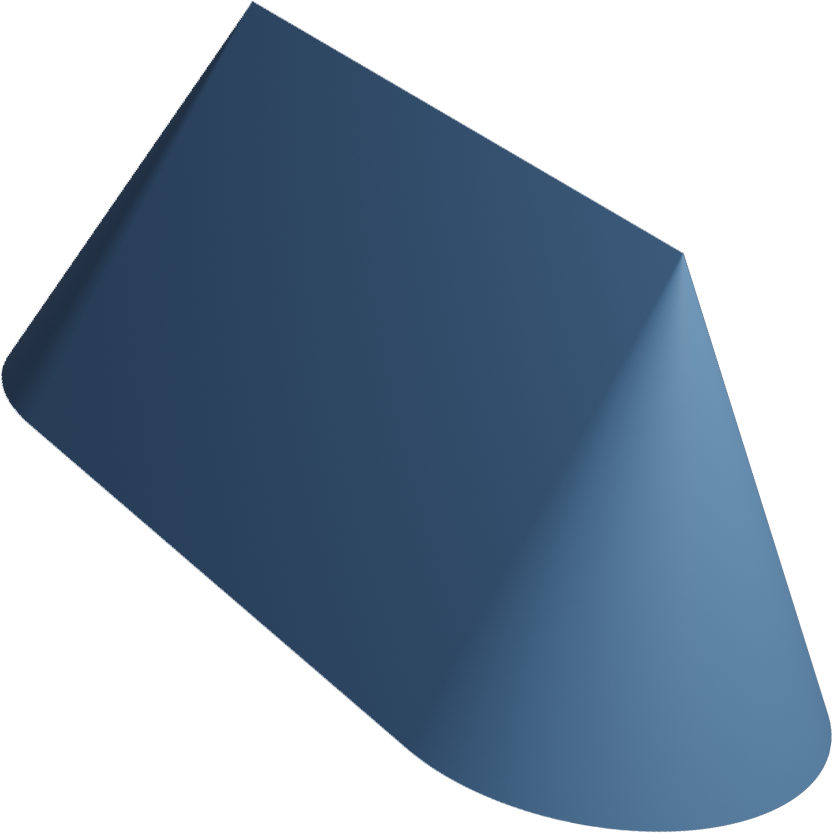}
        \caption{\it 3D Shape of the distance to a single line.}
        \label{renders/0}
    \end{subfigure}
    \hfill
        \begin{subfigure}[t]{0.19\textwidth}
        \includegraphics[width=1.0\textwidth]{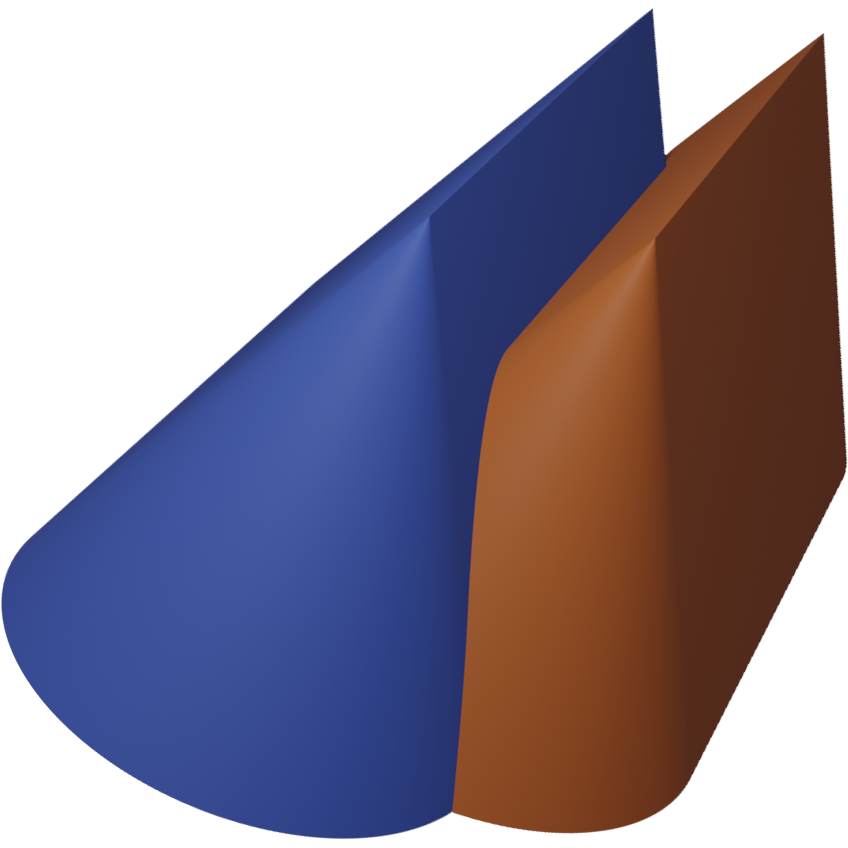}
        \caption{\it Intersection of two 3D distance functions.}
        \label{renders/1}
    \end{subfigure}
        \hfill
        \begin{subfigure}[t]{0.19\textwidth}
        \includegraphics[width=1.0\textwidth]{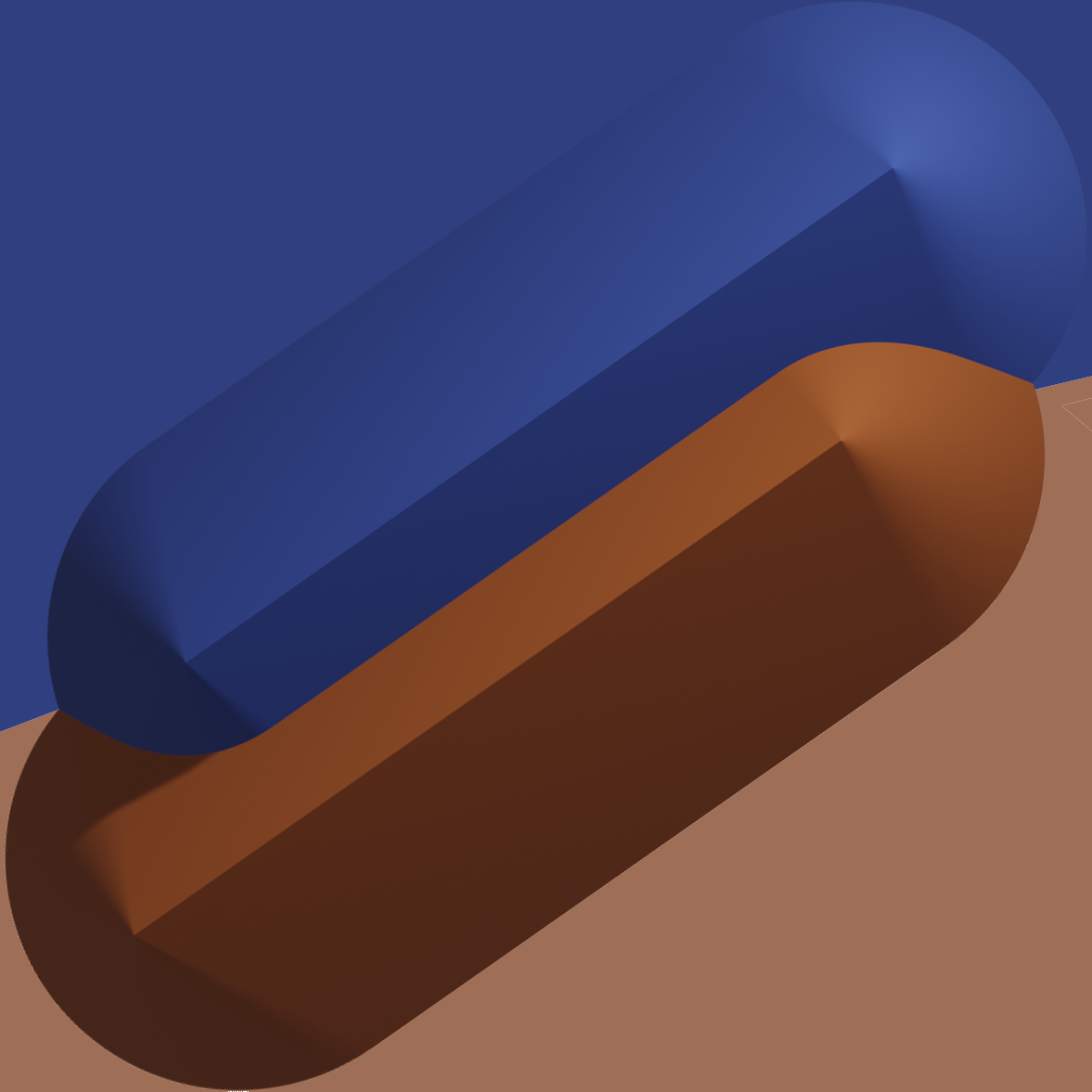}
        \caption{\it The top view of the intersection.}
        \label{renders/3}
    \end{subfigure}
        \begin{subfigure}[t]{0.19\textwidth}
        \includegraphics[width=1.0\textwidth]{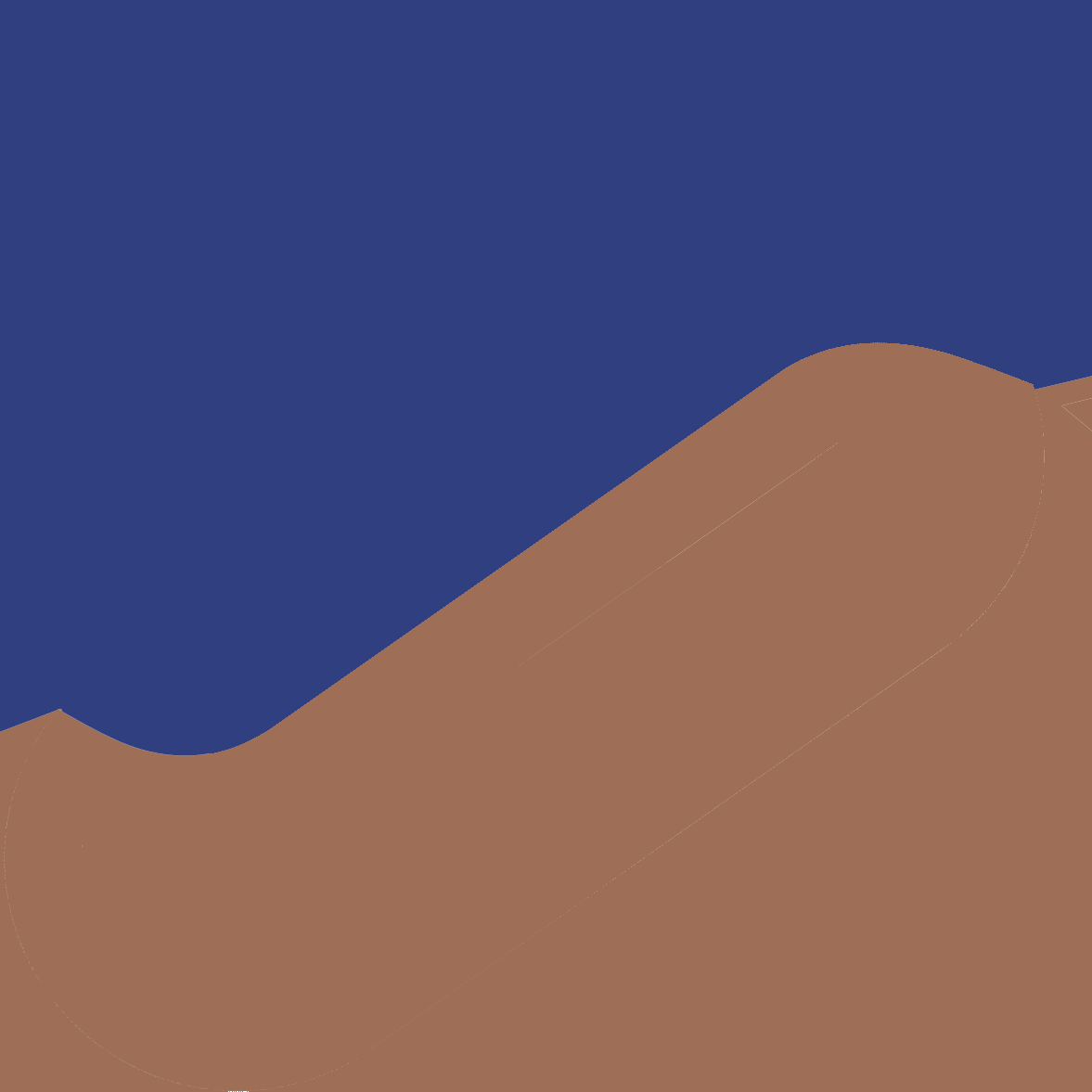}
        \caption{\it  Voronoi partition.}
        \label{renders/4}
    \end{subfigure}
\caption{\it The essence of the method developed by Hoff et al.\cite{keyser1999}. The top view of the intersection of two 3D shapes that come from 3D distance functions directly provides Voronoi partition. }
\label{renders}
\end{figure}

For Voronoi Tessellation with lines as Voronoi sites can be computed in image space by observing that the distance functions can be represented as 3D shapes that consist of cones and triangular prisms as shown in Figure~\ref{renders/0} \cite{keyser1999}. Using this property, Voronoi decomposition of multiple lines can be obtained by creating these shapes in 3D using different colors as shown in Figure~\ref {renders/1} and looking at them from the top view as shown in Figure~\ref{renders/4}. This property makes it easier to implement Voronoi decomposition in graphics hardware using lines as Voronoi sites. Multiple lines and curves can also be implemented using union of the distance to a set of line segments. 

\subsection{Polylines and Curves as Voronoi Sites}

Using only single lines as Voronoi sites, we can produce non-convex tiles as shown in Figures~\ref{Square_Examples}, and~\ref{Hexagonal_Examples}. To construct such complicated tile shapes, one can use multiple lines as a single Voronoi site. As shown in Figure~\ref{Polyline_examples} using multiple lines as Voronoi sites can produce significantly more complicated tiles with curved boundaries. 

\begin{figure}[!htpb]
    \centering
        \begin{subfigure}[t]{0.49\textwidth}
        \includegraphics[width=1.0\textwidth]{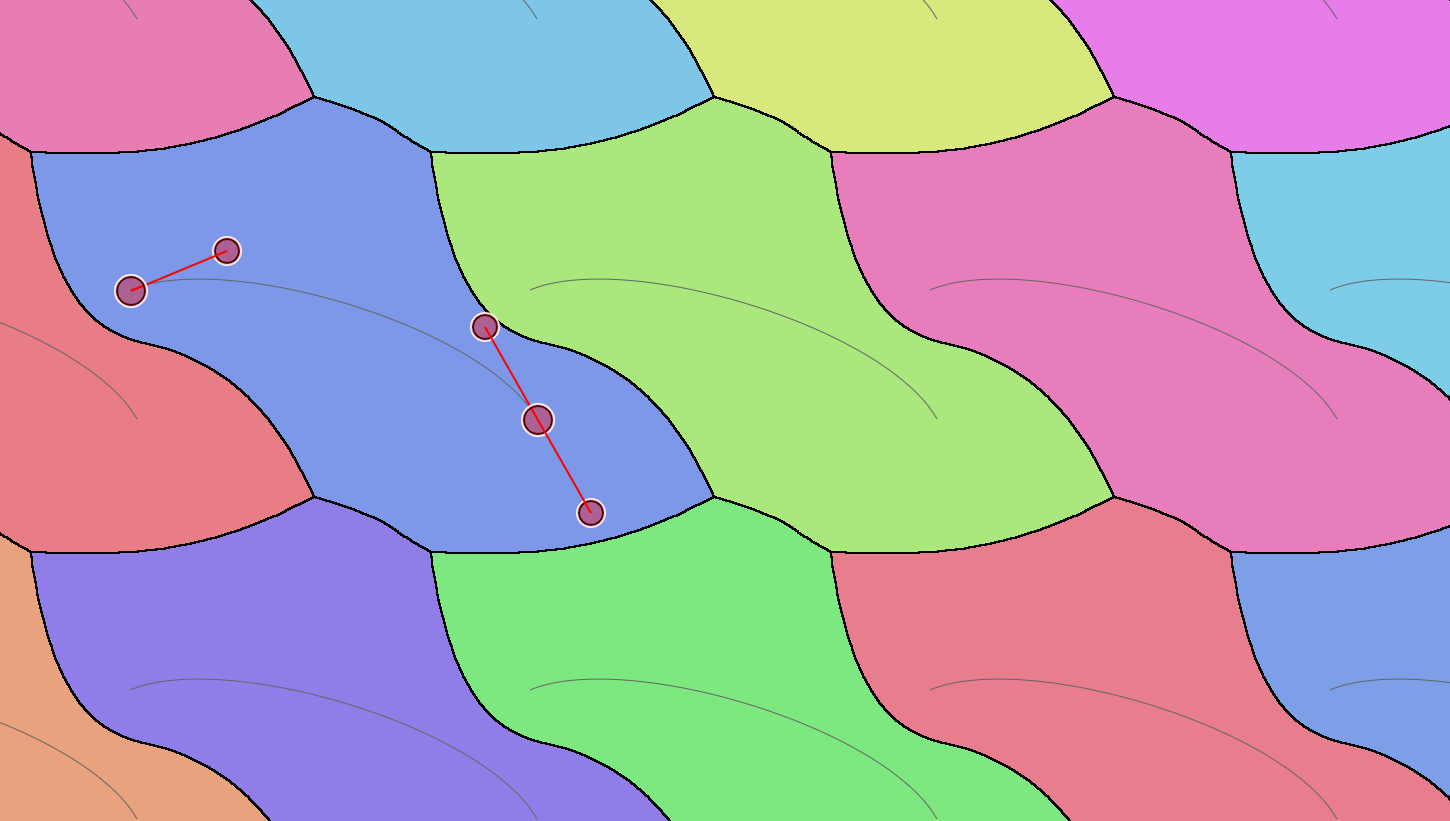}
        \caption{\it Single Cubic curve with p1 symmetry.}
        \label{curves/p1_0}
    \end{subfigure}
    \hfill
        \begin{subfigure}[t]{0.49\textwidth}
        \includegraphics[width=1.0\textwidth]{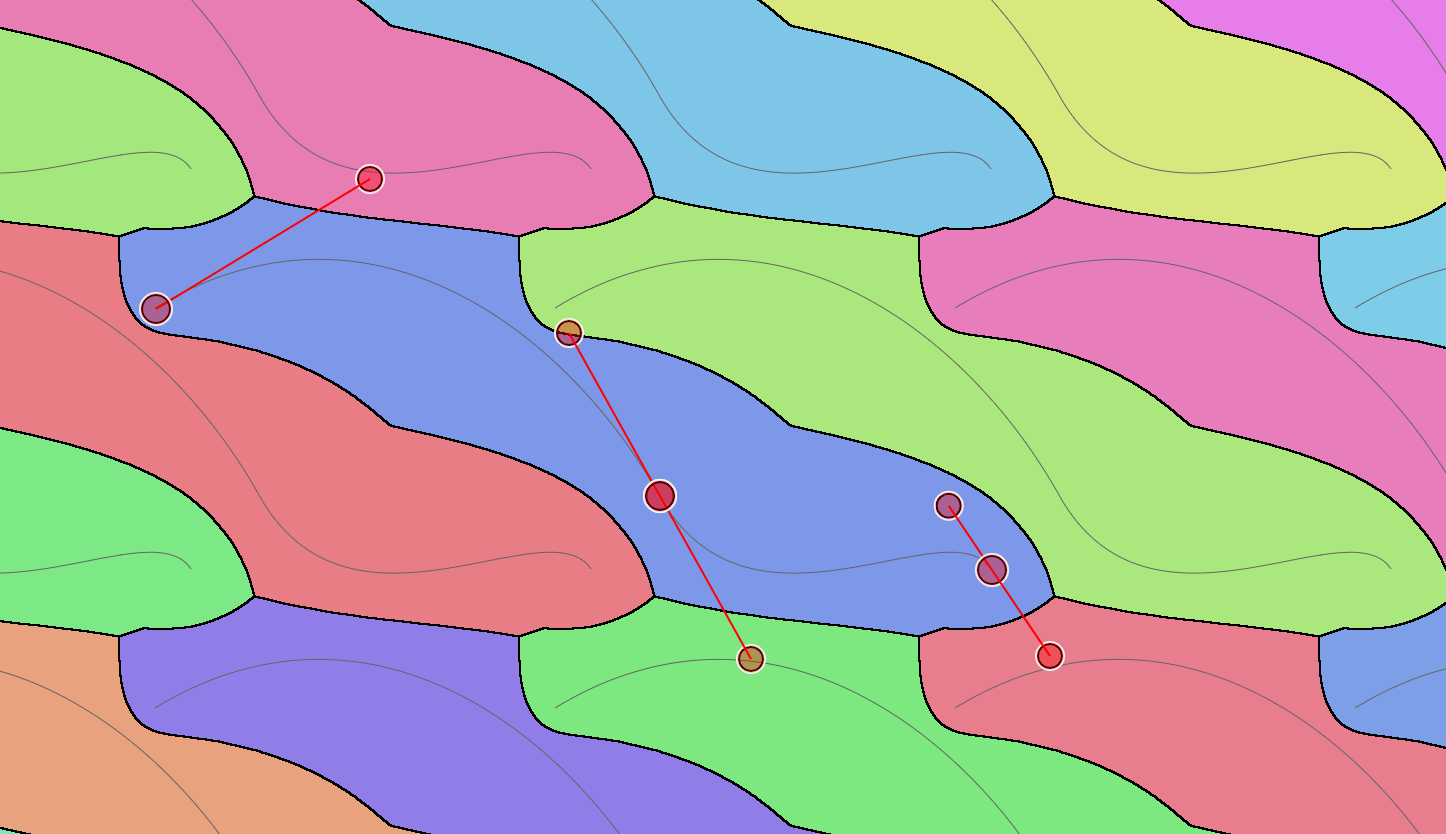}
        \caption{\it Two Cubic curves with p1 symmetry.}
        \label{curves/p1_1}
    \end{subfigure}
\caption{\it Two examples of P1 symmetry by using Hermitian curves.  }
\label{curvesp1}
\end{figure}

\begin{figure}[!htpb]
    \centering
    \begin{subfigure}[t]{0.49\textwidth}
        \includegraphics[width=1.0\textwidth]{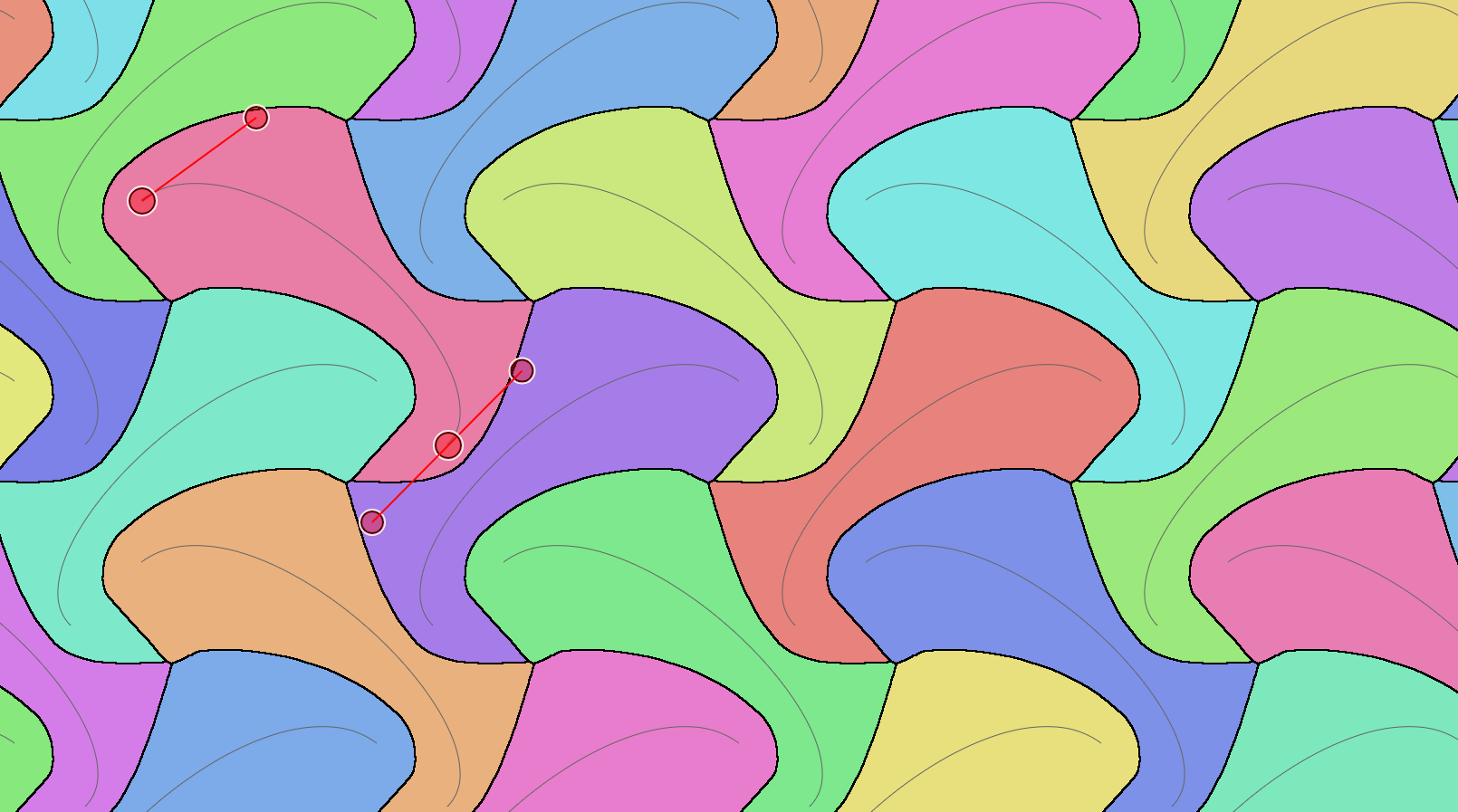}
        \caption{\it Single Cubic curve with pg symmetry.}
        \label{curves/pg_0}
    \end{subfigure}
    \hfill
        \begin{subfigure}[t]{0.49\textwidth}
        \includegraphics[width=1.0\textwidth]{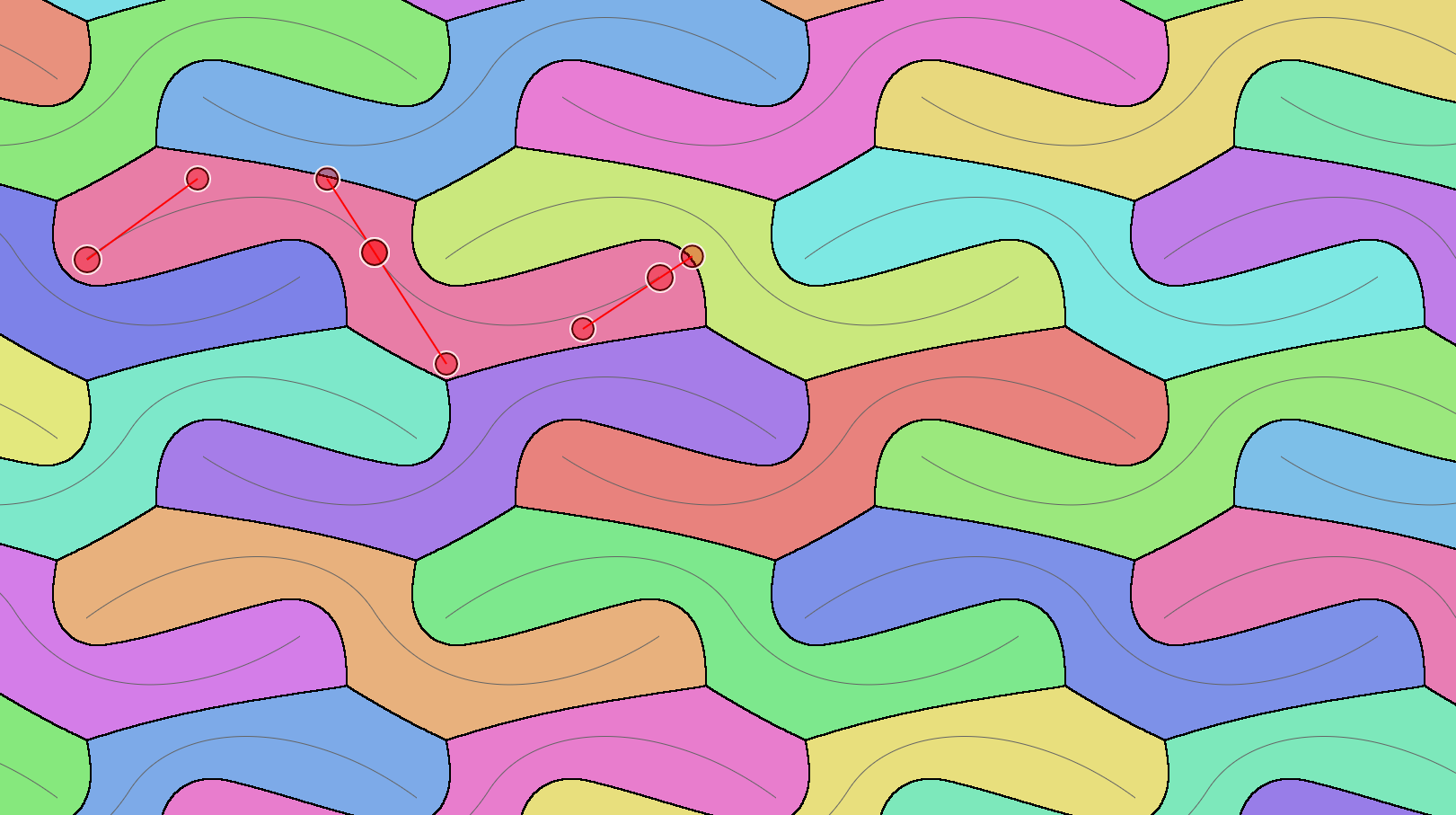}
        \caption{\it Two Cubic curves with pg symmetry.}
        \label{curves/pg_1}
    \end{subfigure}
\caption{\it Two examples of Pg symmetry by using Hermitian curves.  }
\label{curvespg}
\end{figure}

\begin{figure}[!htpb]
    \centering
    \begin{subfigure}[t]{0.49\textwidth}
        \includegraphics[width=1.0\textwidth]{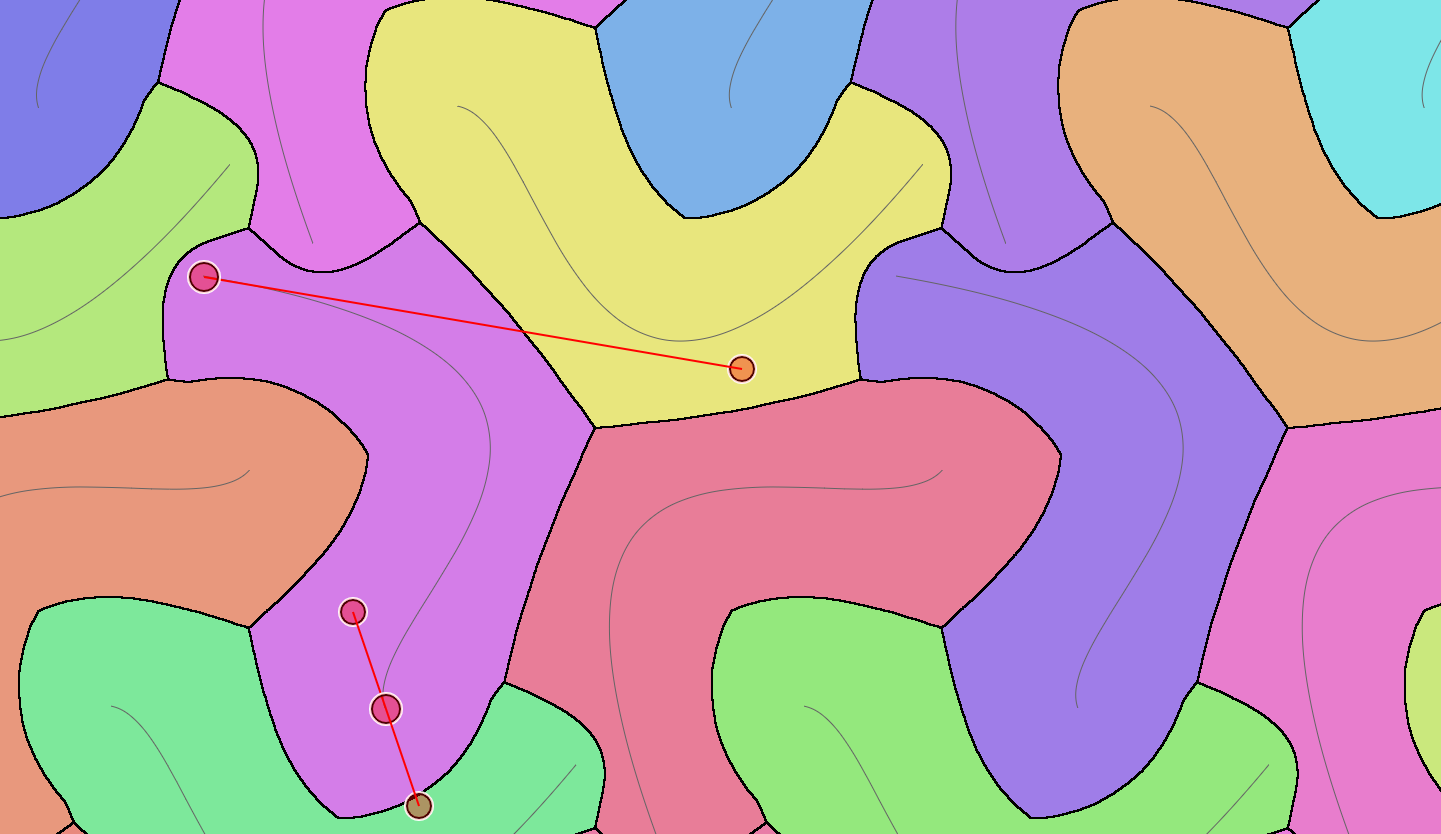}
        \caption{\it Single Cubic curve with p3 symmetry.}
        \label{curves/p3_0}
    \end{subfigure}
    \hfill
        \begin{subfigure}[t]{0.49\textwidth}
        \includegraphics[width=1.0\textwidth]{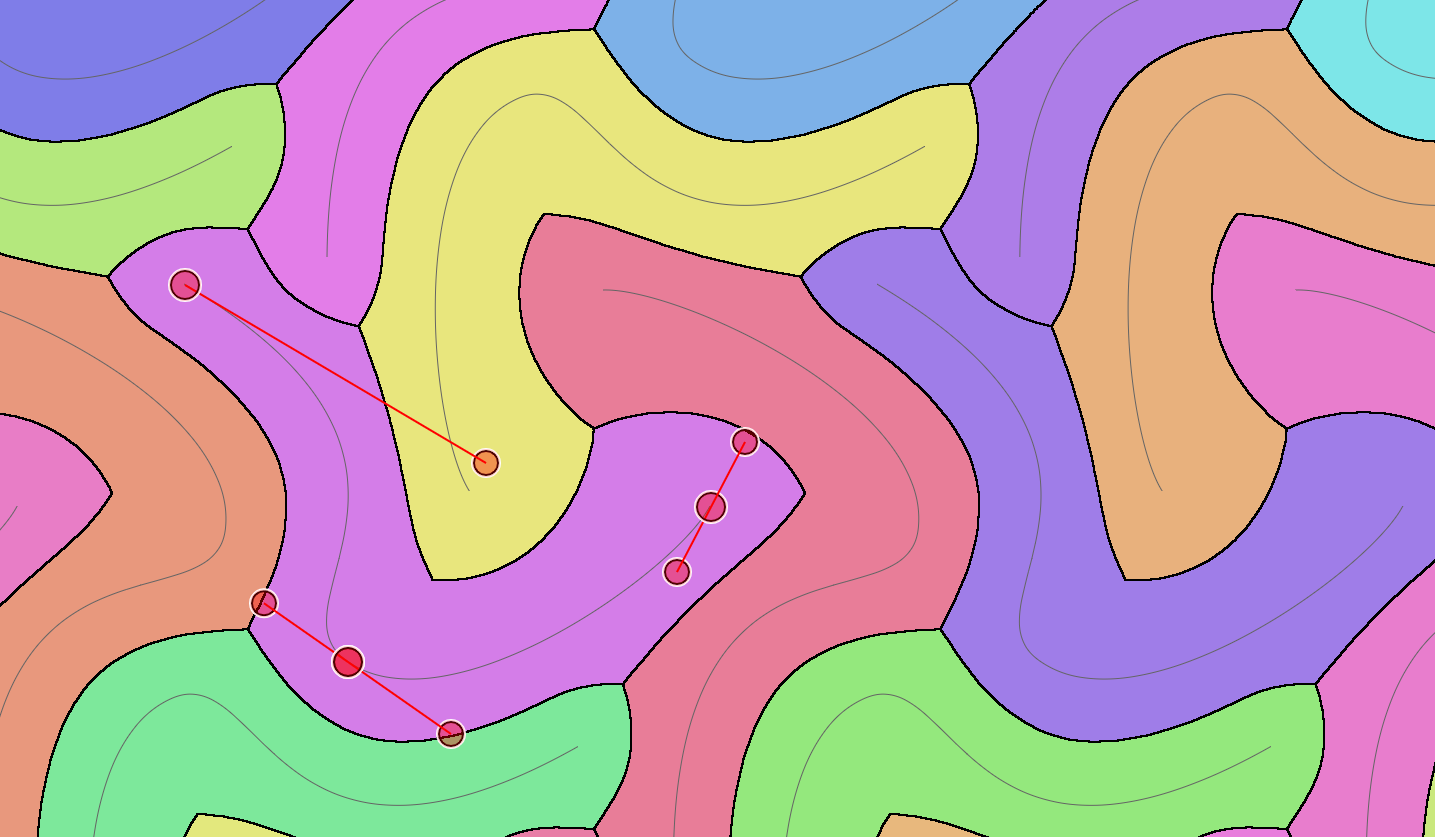}
        \caption{\it Two Cubic curves with p3 symmetry.}
        \label{curves/p3_1}
    \end{subfigure}
\caption{\it Two examples of P3 symmetry by using Hermitian curves.  }
\label{curvesp3}
\end{figure}

\begin{figure}[!htpb]
    \centering
    \begin{subfigure}[t]{0.49\textwidth}
        \includegraphics[width=1.0\textwidth]{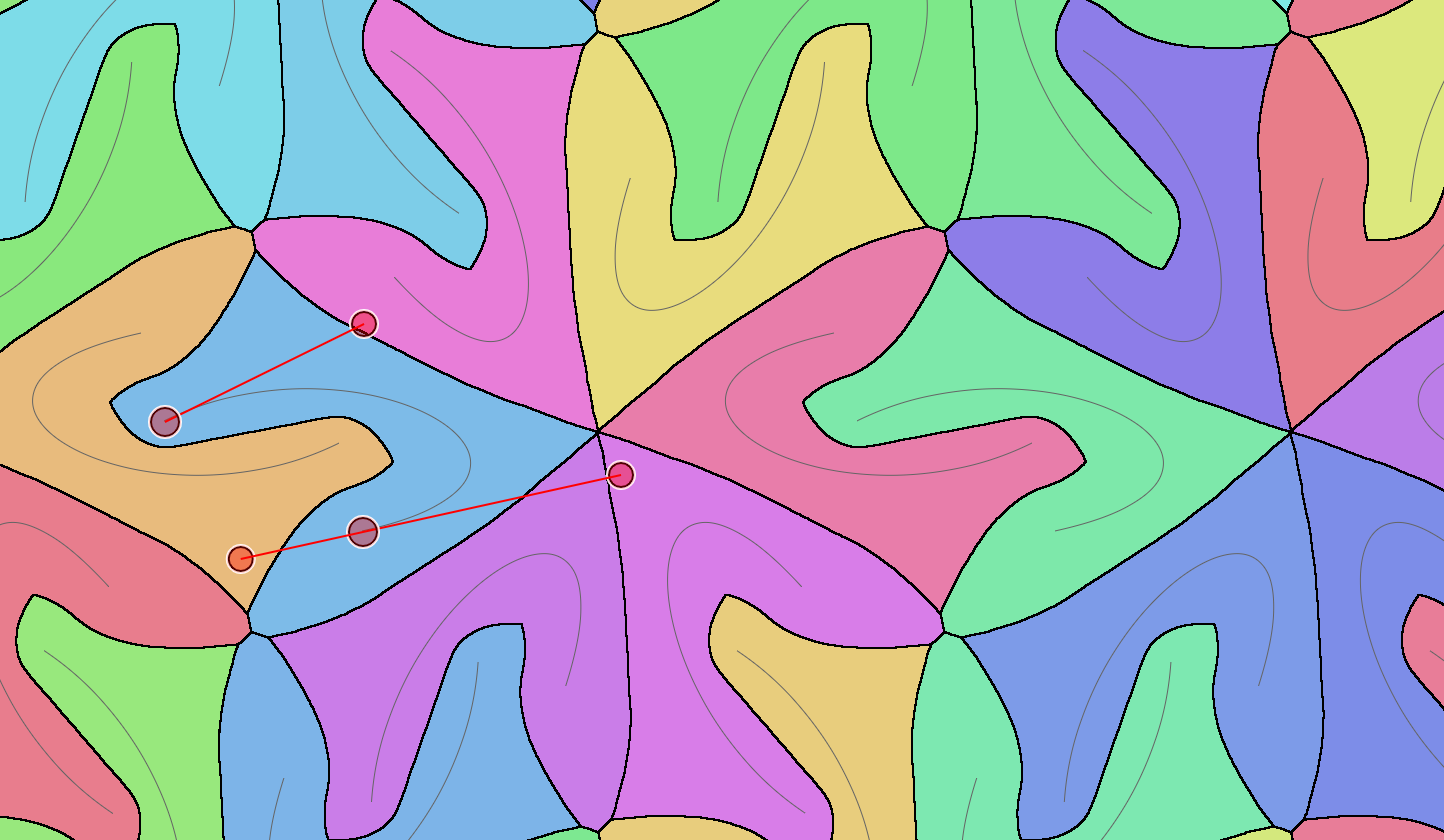}
        \caption{\it Single Cubic curve with p6 symmetry.}
        \label{curves/p6_0}
    \end{subfigure}
    \hfill
        \begin{subfigure}[t]{0.49\textwidth}
        \includegraphics[width=1.0\textwidth]{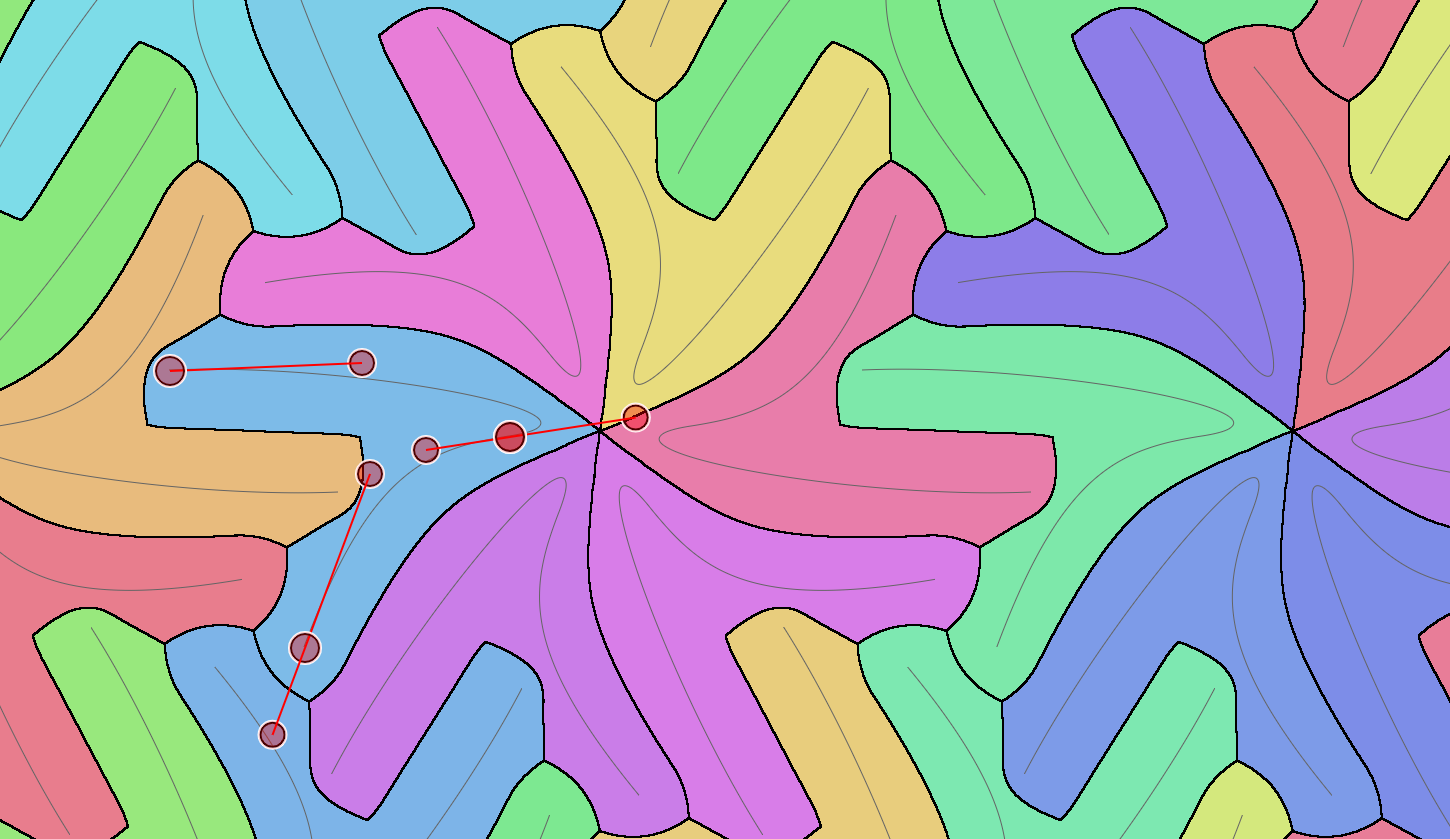}
        \caption{\it Two Cubic curves with p6 symmetry.}
        \label{curves/p6_1}
    \end{subfigure}
\caption{\it Two examples of P6 symmetry by using Hermitian curves.  }
\label{curvesp6}
\end{figure}

The only constraint with multiple lines is that the 
Voronoi sites are still essentially line segments. Therefore, the boundaries can only consist of parabolic curves. To obtain more general boundary curves, it is necessary to extend Voronoi sites to curves. Bezier and B-spline curves \cite{bartels1995introduction} are appropriate since they can efficiently represent a few number of line segments through subdivision processes such as DeCasteljau, Chaikin, and Catmull-Clark subdivision algorithms \cite{boehm1999casteljau,chaikin1974algorithm,catmull1998recursively}. The Figures~\ref{curvesp1},~\ref{curvespg},~\ref{curvesp3},   and~\ref{curvesp6} show examples of curves as Voronoi sites.  

\section{implementation}

For real-time and interactive Voronoi decomposition of 2D space, we have implemented the image-based Voronoi tessellation of 2D space approach developed by Hoff et al.\cite{keyser1999} using graphics hardware. This approach utilizes the efficiency of the 3D graphics pipeline for discrete Voronoi diagram computation on a rectangular grid of points (pixels). 

For each primitive, we provide the distance function in the form of a 3D polygonal mesh \cite{keyser1999}. For a point primitive 3D polygonal mesh is a cone and for a line primitive the polygonal mesh consists of a "tent" with a cone at each of the two ends (See Figure~\ref{renders}). Our system consists of an effective combination of three existing methods: (1) The Voronoi tessellation of 2D space using graphics hardware; (2) Symmetry operations over the Voronoi sites; and (3) Coloring and Detecting Boundaries. 

These polygonal meshes are rendered at the position of each primitive. This process defines the Voronoi region for each primitive. The scene is rendered using a polygon scan conversion and Z-buffer depth comparison engine. The resulting top view of the scene is an image that is the Voronoi tessellation of the 2D space. 

We have implemented a web-based version and it is available at   \href{https://voronoi.viz.tamu.edu/}{https://voronoi.viz.tamu.edu/}. In this web-based version, we used the open-source code repository, EscherSketch \cite{levskaya2017}, to generate the seventeen planar symmetries for implementing symmetry operations over the Voronoi sites. We also used the implementation of the drawing interface from the repository, with minor changes. In addition to the web-based version, we also implemented this system as a plug-in in Blender, mostly using Blender functions. 

For coloring and detecting boundaries (in both versions), we assign a unique color to each region of the Voronoi tessellation by computing pseudo-random values in the HSV color space to avoid two neighboring regions being assigned the same color. The discrete boundaries of the regions are computed using the Sobel Operator to detect edges \cite{Kanopoulos1988-ki} in the rendered image as the change in color would imply a change in the Voronoi region \cite{keyser1999}.

We have also implemented curves to obtain more complicated boundaries beyond parabolas. For a simple interface, we use the Hermitian form. Users define positions and tangents to design curves. We then transform these curves into the third-degree Bezier polynomials. The advantage of the Bezier form, the curves can be constructed by using the DeCasteljau subdivision algorithm. The main advantage of the DeCasteljau subdivision is that it gives a good approximation of the curve only with 12 or 24 line segments.

To simplify the curve description for users, we also used the Catmull-Rom formulation that allows to use only positions of points to describe Hermitian curves by implicitly computing tangent vectors \cite{catmull1974class,yuksel2011parameterization}. Since the Catmull-Rom curves are also piecewise cubic curves, it is also straightforward to transform the Catmull-Rom curves into the third-degree Bezier polynomials.

\section{Conclusion and Future Work}

In this paper, we present an approach for the development of real-time systems to interactively design symmetric tiles with curved edges. Our approach is an extended version of the image-based Voronoi decomposition method \cite{keyser1999}. We made two extensions: (1) We use higher-order shapes such as curves as Voronoi sites instead of just points and line segments, and (2)  we use Voronoi sites that are closed under wallpaper symmetries. These two extensions allow us to create Voronoi tessellations with 2D space-filling tiles with curved edges. We have developed both a web application and a stand-alone application that can allow the design of space-filling tiles with curved edges interactively in real time. 

We have also demonstrated that using Voronoi-based methods not all symmetry operations are useful for creating curved tiles: all symmetries that use mirror operation produce straight lines that are useless for creating new tiles. This result is interesting because it suggests that we need to avoid mirror transformations to produce unusual space-filling tiles in 2D and 3D using Voronoi tessellations.

This work can be extended for the construction of 2.5D space-filling structures such as Delaunay Lofts 
\cite{subramanian2019delaunay}, Generalized Abeille Tiles \cite{akleman2020generalized}, and VoroNoodles \cite{ebertvoronoodles,mullins2022voronoi}. We currently work on the development of interactive systems to provide a generalization of all of these methods as layer-by-layer curves. 

\subsection{Individual Contributions}

Haard Panchal implemented the web-based system. Ergun Akleman and Vinayak Krishnamurthy developed the main approach in this paper. The paper is written mostly Ergun Akleman with initial text by Haard Panchal. Tolga Talha Yildiz implemented a Blender-based interactive system. Varda Grover implemented Catmull-Rom curve interfaces into web-based system. Varda Grover and Tolga Talha Yildiz are currently implementing 2.5D systems for web-based and stand-alone systems. 

\bibliographystyle{unsrtnat}
\bibliography{references}
\end{document}